\documentclass{emulateapj}
\shortauthors{Gabel et al.}
\shorttitle{Global Fitting of Intrinsic UV Absorption in Mrk 279}

\submitted{Accepted for publication in ApJ}

\begin{document}
\title{X-ray/UV Observing Campaign on the Mrk 279 AGN Outflow:\\
A Global Fitting Analysis of the UV Absorption \altaffilmark{1}}
\author{Jack R. Gabel\altaffilmark{2}, Nahum Arav\altaffilmark{2}, 
Jelle S. Kaastra\altaffilmark{3}, Gerard A. Kriss\altaffilmark{4,5},
Ehud Behar\altaffilmark{6}, Elisa Costantini\altaffilmark{3,7}, 
C.~Martin Gaskell\altaffilmark{8}, 
Kirk T. Korista\altaffilmark{9}, Ari Laor\altaffilmark{6}, 
Frits Paerels\altaffilmark{10}, Daniel Proga\altaffilmark{11},
Jessica Kim Quijano\altaffilmark{4}, 
Masao Sako\altaffilmark{12}, Jennifer E. Scott\altaffilmark{4}, 
Katrien C. Steenbrugge \altaffilmark{3}}
\altaffiltext{1}{Based on observations made with the NASA/ESA {\it Hubble 
Space Telescope} and the NASA-CNES-CSA {\it Far Ultraviolet Spectroscopic Explorer},
and obtained at the Space Telescope Science Institute, which is operated by the 
Association of Universities for Research in Astronomy, Inc. under NASA 
contract NAS~5-26555.}
\altaffiltext{2}{Center for Astrophysics and Space Astronomy, 
University of Colorado, 389 UCB, Boulder CO 80309-0389; 
jgabel@colorado.edu, arav@colorado.edu}
\altaffiltext{3}{SRON National Institute for Space Research, Sorbonnelaan 2,
3584 CA Utrecht, The Netherlands; J.S.Kaastra@sron.nl, e.costantini.sron.nl, 
K.C.Steenbrugge@sron.nl}
\altaffiltext{4}{Space Telescope Science Institute, 3700 San Martin Drive, 
Baltimore, MD 21218; gak@stsci.edu; jescott@stsci.edu; jkim@stsci.edu}
\altaffiltext{5}{Center for Astrophysical Sciences, Department of Physics and 
Astronomy, The Johns Hopkins University, Baltimore, MD 21218}
\altaffiltext{6}{Department of Physics, Technion, Haifa 32000, Israel;
behar@physics.technion.ac.il, laor@physics.technion.ac.il}
\altaffiltext{7}{Astronomical Institute, University of Utrecht, PO Box 80 000,
3508 TA Utrecht, The Netherlands}
\altaffiltext{8}{Department of Physics and Astronomy, University of Nebraska,
Lincoln NE 68588-0111; mgaskell1@unl.edu}
\altaffiltext{9}{Department of Physics, Western Michigan University, 
Kalamazoo, MI 49008; korista@wmich.edu}
\altaffiltext{10}{Columbia Astrophysics Laboratory, 550 West 120th Street,
New York, NY 10027; frits@astro.columbia.edu}
\altaffiltext{11}{JILA, University of Colorado, Campus Box 440, Boulder, CO
80309; proga@colorado.edu}
\altaffiltext{12}{Theoretical Astrophysics and Space Radiation Laboratory,
California Institute of Technology, MC 130-33, Pasadena, CA 91125; 
masao@tapir.caltech.edu}

\begin{abstract}
  We present an analysis of the intrinsic UV absorption in the Seyfert 1
galaxy Mrk 279 based on simultaneous long observations with the {\it Hubble 
Space Telescope} (41 ks) and the {\it Far Ultraviolet Spectroscopic 
Explorer} (91 ks).
To extract the line-of-sight covering factors and ionic column densities, 
we separately fit two groups of absorption lines:
the Lyman series and the CNO lithium-like doublets.
For the CNO doublets we assume that all three ions share the same
covering factors.  The fitting method applied here overcomes some limitations
of the traditional method using individual doublet pairs; it allows
for the treatment of more complex, physically realistic scenarios
for the absorption-emission geometry and eliminates systematic
errors that we show are introduced by spectral noise.
We derive velocity-dependent solutions based on two
models of geometrical covering -- a single covering factor for 
all background emission sources, and separate covering factors for the
continuum and emission lines.  
Although both models give good statistical fits to the
observed absorption, we favor the model with two covering factors 
because:
(a) the best-fit covering factors for both emission sources
are similar for the independent Lyman series and CNO doublet fits;
(b) the fits are consistent with full coverage of the 
continuum source and partial coverage of the emission lines by the absorbers,
as expected from the relative sizes of the nuclear emission components; and
(c) it provides a natural explanation for variability in the 
Ly$\alpha$ absorption detected in an earlier epoch.  
We also explore physical and geometrical constraints on the outflow 
from these results.
\end{abstract}

\keywords{galaxies: individual (Mrk 279) --- galaxies: active --- galaxies: Seyfert --- ultraviolet: galaxies}

\section{Introduction}  

     Mass outflow, seen as blueshifted absorption in UV and X-ray spectra,
is an important component of active galactic nuclei \citep[AGNs; see recent
review in][]{cren03}.
This ``intrinsic absorption" is ubiquitous in nearby AGNs, appearing
in over half of Seyfert 1 galaxies having high-quality UV spectra obtained 
with the {\it Hubble Space Telescope} 
\citep[{\it HST};][]{cren99} and the {\it Far Ultraviolet Spectroscopic Explorer} 
\citep[{\it FUSE};][]{kris02}.  
Spectra from the {\it Advanced 
Satellite for Cosmology and Astrophysics (ASCA)} identified X-ray ``warm absorbers", 
seen as absorption edges, in a similar percentage of objects \citep{reyn97,geor98}.  
Large total ejected masses have been inferred for these 
outflows, exceeding the accretion rate of the central black hole in some cases, 
indicating mass outflow plays an important role in the overall energetics in 
AGNs \citep[e.g.][]{reyn97}.
Recent studies have recognized and explored the 
potential effect of outflows on all scales of the AGN environment, from feeding 
the central supermassive black hole in AGNs \citep{blan99,blan04}, to 
influencing the evolution of the host galaxy \citep{silk98,scan04} 
and the metallicity of the intergalactic medium \citep{cava02}.

    Measured ionic column densities provide the basis for 
interpretation of the physical nature of AGN outflows.
Detailed UV spectral studies over the past decade have shown measurements of these crucial 
parameters are often not straightforward.
Analyses of absorption doublets and multiplets has revealed the absorbers
typically only partially occult the background emission sources.
Without proper treatment of this effect, the column densities 
could be severely in error \citep[e.g.][]{wamp93,barl97,hama97}.  
Additional complications that could affect column density measurements
are different covering factors for different background emission sources 
\citep{gang99,gabe03}, velocity-dependent covering factors \citep[e.g.][]{arav99}, 
and inhomogeneous distributions of absorbing material \citep{deko02}.

     Many recent investigations of AGN outflows have focused on intensive
multiwavelength observations of Seyfert 1 galaxies.
Seyferts are well suited for these studies because they include the brightest 
AGNs in the UV and X-ray.
The X-ray spectra contain the imprint of the bulk of the outflow's mass,
which can now be deblended into individual absorption lines 
with the high-resolution capabilities of the 
{\it Chandra X-ray Observatory (CXO)} and {\it XMM-Newton Space Observatory},
allowing detailed study.
The high quality UV spectra available with {\it HST} and {\it FUSE}
provide a complimentary, precise probe of the complex absorption troughs.
Due to the relatively narrow absorption in Seyfert outflows,
the important UV doublets and multiplets are typically unblended, allowing
measurements of these key diagnostic lines.

     We have undertaken an intensive multiwavelength observing
campaign with {\it HST}/STIS, {\it FUSE}, and {\it CXO} to study the 
intrinsic absorption in the Seyfert 1 galaxy Mrk 279.  
Mrk 279 was selected for this study because of its 
UV and X-ray brightness and the rich absorption spectrum in both bands, including 
unblended and well-resolved UV doublets \citep[see][hereafter SK04]{scot04}.
Additionally, it has minimal contamination by Galactic absorption and 
a relatively weak contribution from a narrow emission line region (NLR), both of
which can complicate measurements of the absorption properties.
As part of a series of papers devoted to this campaign, we present here
a detailed study of the UV absorption in the combined STIS and {\it FUSE} spectra.
We develop a new approach for measuring the covering factors and column densities
in the absorbers, making full use of the high quality far-UV spectrum.
These measurements provide the foundation for subsequent analysis and interpretation
of the mass outflow in Mrk 279, and provide novel geometric constraints.
In parallel papers, we present analysis of the X-ray spectrum \citep{cost04}, 
inhomogeneous models of the UV absorption \citep{arav04}, and density diagnostics
based on O\,{\sc v} K-shell X-ray lines \citep{kaas04}.
In future papers, we will present photoionization
models of the UV and X-ray absorption and analysis of absorption variability.
In the next section, we describe the {\it HST} and {\it FUSE} observations 
and present an overview of the absorption spectrum; 
in \S 3, we review the standard doublet technique for measuring intrinsic
absorption and, together with an Appendix, discuss important limitations of this method; 
the formalism of our fitting method and results for Mrk 279 are described in \S 4;
in \S 5, the fits are interpreted, and implications for physical constraints on the 
outflow are explored; finally, a summary is presented in \S 6.

\section{Observations and the Intrinsic Absorption Spectrum}

\subsection{Simultaneous {\it HST}/STIS and {\it FUSE} Observations of Mrk 279}

  The nucleus of Mrk 279 was observed for a total of 41 ks (16 orbits) with the Space
Telescope Imaging Spectrograph (STIS) on board {\it HST} between 
2003 May 13 -- 18 and for 91 ks with {\it FUSE} between 2003 May 12 -- 14. 
The STIS observation used the E140M grating, which covers 
1150 -- 1730 \AA, and was obtained through the 0$\arcsec$.2 $\times$ 0$\arcsec$.2 aperture.
The spectrum was processed with CALSTIS v2.16, which removes the background light
from each echelle order using the scattered light model from \citet{lind00}.
Low residual fluxes in the cores of saturated Galactic lines indicate accurate
removal of scattered light: typical fluxes in the cores are $\pm <$ 2.5\% of the
local unabsorbed continuum flux levels, and mean fluxes averaged over the absorption
cores are $<$ 3\% of the noise in the troughs.
The final spectrum was sampled in 0.012 -- 0.017 \AA\ bins, thereby preserving the
$\sim$ 6.5~km~s$^{-1}$ kinematic resolution of STIS/E140M.

    We found that the standard pipeline processing did not yield a fully
calibrated spectrum due to two effects: a) the echelle ripple structure, due to 
the characteristic efficiency of the detector along each order, is
not completely removed \citep[see][]{heap97}, and b) the sensitivity of
the MAMA detectors has degraded with time, and the change has not been
incorporated in the pipeline for the echelle gratings. In order to correct
for these effects in the Mrk~279 spectrum, we used multiple spectra of the
white dwarf spectrophotometric standard, BD+28~4122, one of which was
taken close in time to our observation.
First, a composite stellar spectrum of BD+28~4122 composed of FOS and STIS
data \citep{bohl01} was used to flux-calibrate a
1997 STIS spectrum of BD+28~4122 that does not exhibit the echelle ripple
structure. This flux-calibrated spectrum was then used to correct a STIS
spectrum of BD+28~4122, taken on 3 May 2003 with the same grating and
aperture as our Mrk~279 observation, and which does show the same ripple
structure seen in the Mrk~279 spectrum.  These two corrections were
performed by dividing the fiducial spectrum for each step by the
comparison spectrum, fitting a polynomial to the result, and then
multiplying the comparison spectrum by that polynomial.
We used the polynomials from the second step described above to correct
the spectrum of Mrk~279.  To obtain a smooth correction for each order in
the Mrk~279 spectrum, we applied averages of the polynomials
corresponding to the four adjacent orders.  Although we were able to
remove most of the echelle ripple structure in this way, some lower
amplitude residual curvature remains in some orders. However, the
intrinsic absorption features measured in this study are well-corrected.


\begin{deluxetable*}{lrr}
\tablecaption{Kinematics of Absorption Components in Mrk 279\tablenotemark{a} \label{tbl-1}}
\tablehead{
\colhead{Component} & \colhead{Velocity}\tablenotemark{b} & \colhead{FWHM} \\ 
\colhead{} & \colhead{km s$^{-1}$} & \colhead{km s$^{-1}$}}
\startdata
1 & 85 & 40 \\
2 & -265 & 50 \\
2a & -290 & 30 \\
2b & -325  & 30 \\
2c & -355 & 65 \\
3 & -390 & 20 \\
4 & -460 & 20 \\
4a & -490 & 65 \\
5 & -550 & 30 \\
\enddata
\tablenotetext{a}{Measurements for components 2, 2a, 2c, and 4a are from the N V $\lambda$1242
line; components 1, 2b, 3, 4, and 5 are from Si III and C III.}
\tablenotetext{b}{Radial velocity relative to the systemic redshift adopted for Mrk 279,
$z = $0.305}
\end{deluxetable*}

   The {\it FUSE} spectrum, obtained through the 30$\arcsec \times$ 30$\arcsec$ aperture,
covers 905 -- 1187 \AA .  The spectrum was processed with the current standard 
calibration pipeline, CALFUSE v2.2.3.  The eight individual spectra obtained with
{\it FUSE}, from the combination of four mirror/grating channels and two 
detectors, were coadded for all exposures.  Mean residual fluxes measured in the cores of 
saturated Galactic lines are consistent with zero within the noise (i.e., standard 
deviation of the fluxes) in the troughs of these lines, indicating accurate background removal.
The spectrum was resampled into $\sim$ 0.02 -- 0.03 \AA\ bins to increase the signal-to-noise 
ratio (S/N) and preserve the full resolution of {\it FUSE}, which is nominally 
$\sim$ 20~km~s$^{-1}$.

   To place the {\it FUSE} and STIS spectra on the proper wavelength scale, 
we followed the procedure described in SK04.  The centroids of prominent, unblended 
Galactic interstellar absorption lines were measured and used as fiducials
in comparing to the Galactic 21 cm H\,{\sc i} line in the line-of-sight to Mrk 279
\citep{wakk01}.  The lines measured in the STIS spectrum are consistent with 
the 21 cm H\,{\sc i} line within the measurement uncertainties and thus required no correction.
The {\it FUSE} spectrum showed substantial shifts relative to the adopted standard. 
Due to non-linear offsets in the wavelength scale, local shifts were measured and applied 
individually to each spectral region containing intrinsic absorption features.

     To normalize the absorption, we fit the total intrinsic (i.e., unabsorbed) 
AGN emission in the {\it FUSE} and STIS spectra over each intrinsic absorption
feature.  This was done empirically by fitting cubic splines to unabsorbed 
spectral regions adjacent to the features, at intervals of $\sim$ 5 \AA\ . 
We also derived models for the individual contributions of the different emission 
sources (continuum and emission lines), since they are required for our analysis.
For the continuum source, we fit a single power law ($f_{\lambda} \propto \lambda^{-\alpha}$) 
to the observed flux at two widely separated wavelengths that are relatively uncontaminated with
absorption or line emission features, $\lambda =$ 955 and 1500 \AA .  
After first correcting for Galactic extinction 
\citep[using the extinction law of][with $E(B-V)=0.016$]{card89}, we find a best-fit spectral
index $\alpha =$ 1.6.
This power law model matches the few other line-free regions of the UV spectrum, 
i.e. $\lambda \approx$ 1150 -- 1180, 1330, and 1390 \AA , to within a few percent.
Thus, for the emission line model, we simply subtracted the continuum power law model from the 
empirical fit to the total emission.

\subsection{The Far-UV Absorption Spectrum}

     The full far-UV spectrum from our {\it FUSE} and STIS observations is shown in Figure 1.  
The active nucleus in Mrk 279 was in a relatively high flux state during this epoch; 
the UV continuum flux was similar to a 2000 January {\it FUSE} 
observation, and $\approx$ 7 times stronger than in {\it FUSE} and STIS spectra
obtained in 2002 May.  Full treatment of these earlier observations is given in SK04.
Qualitatively, the intrinsic absorption spectrum in our new observations
is similar to the earlier epochs (although some important variations were detected 
that will be the subject of a later study).
Here, we give a brief overview, and refer the reader to SK04 for a 
more thorough phenomenological discussion of the absorption.

\begin{figure*}[location=t]
\includegraphics[angle=90,width=17cm]{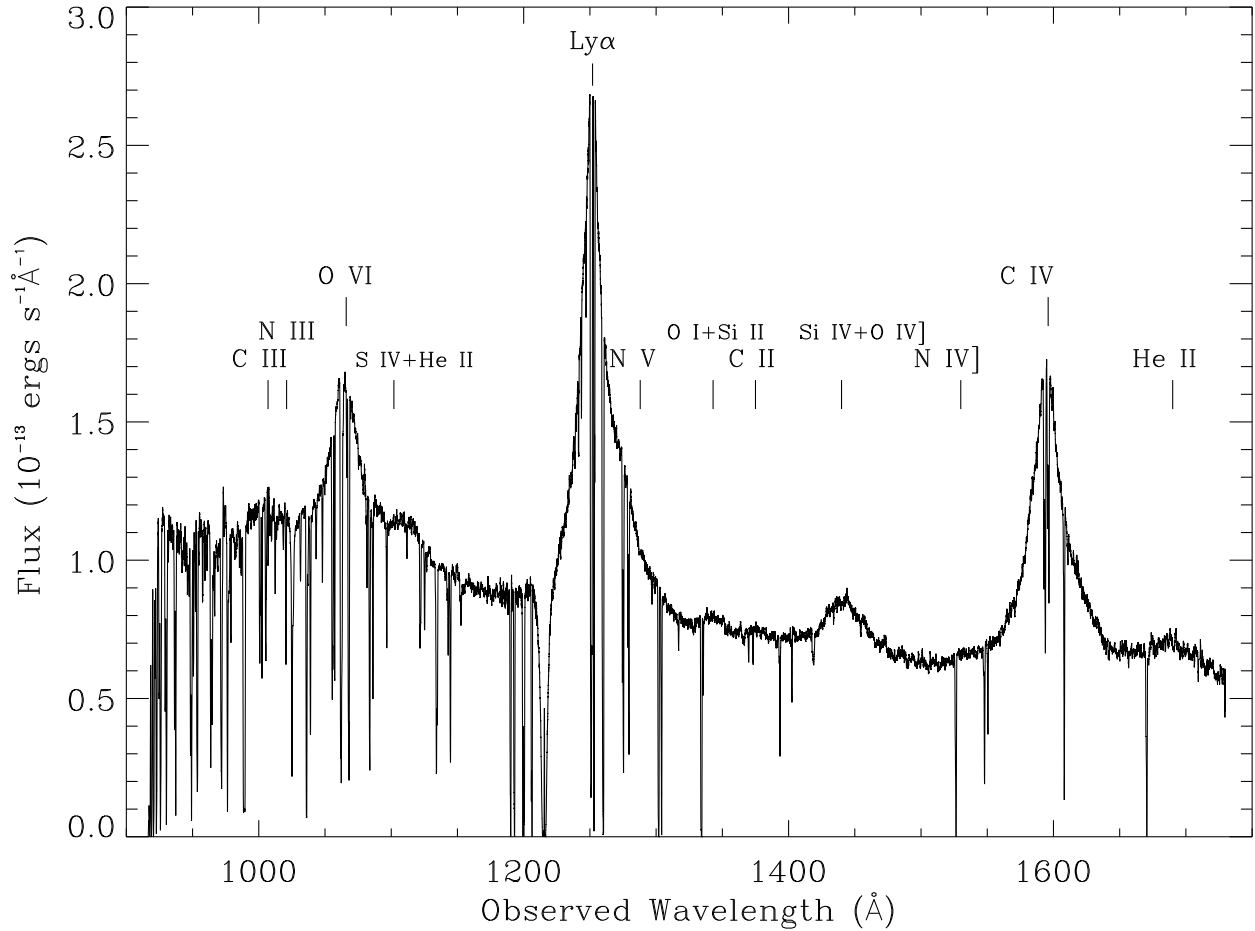}
\vspace*{0.1 in}
\caption{Full far-UV spectrum of the active nucleus in Mrk 279 from the {\it FUSE} and
{\it HST}/STIS spectra obtained in May 2003.  Emission lines are labeled above
the spectrum.  The data were heavily binned for clarity in presentation. \label{fig1}}
\end{figure*}

   Absorption from a range of ions is detected 
around the systemic velocity of the host galaxy, between $v = -$600 to $+$150 km~s$^{-1}$; 
we adopt the redshift for Mrk 279 from SK04, $z =$ 0.0305 $\pm$ 0.0003.
Normalized absorption profiles for some of the prominent lines are shown in Figure 2.
The absorption is seen to be resolved into multiple distinct kinematic components at the
resolution of STIS E140M and {\it FUSE}, revealing striking differences 
in the kinematic structure of different ions.  
Low-ionization species appear in several narrow components 
(see Si\,{\sc iii} in Figure 2, but also Si\,{\sc ii}, C\,{\sc ii}, C\,{\sc iii}, 
and N\,{\sc iii} in SK04 Figures 7 -- 14).
However, the more highly ionized O\,{\sc vi}, N\,{\sc v}, and C\,{\sc iv} doublets, which are 
the primary UV signatures of intrinsic absorption in AGNs, are much broader and have different
centroid velocities.  The Lyman lines exhibit the kinematic structure of the low-ionization
components, but also appear in the lower outflow velocity region coinciding with
the high-ionization lines, $v \approx -$300 to $-$200 km s$^{-1}$.

\begin{figure}
\includegraphics[width=8.5cm]{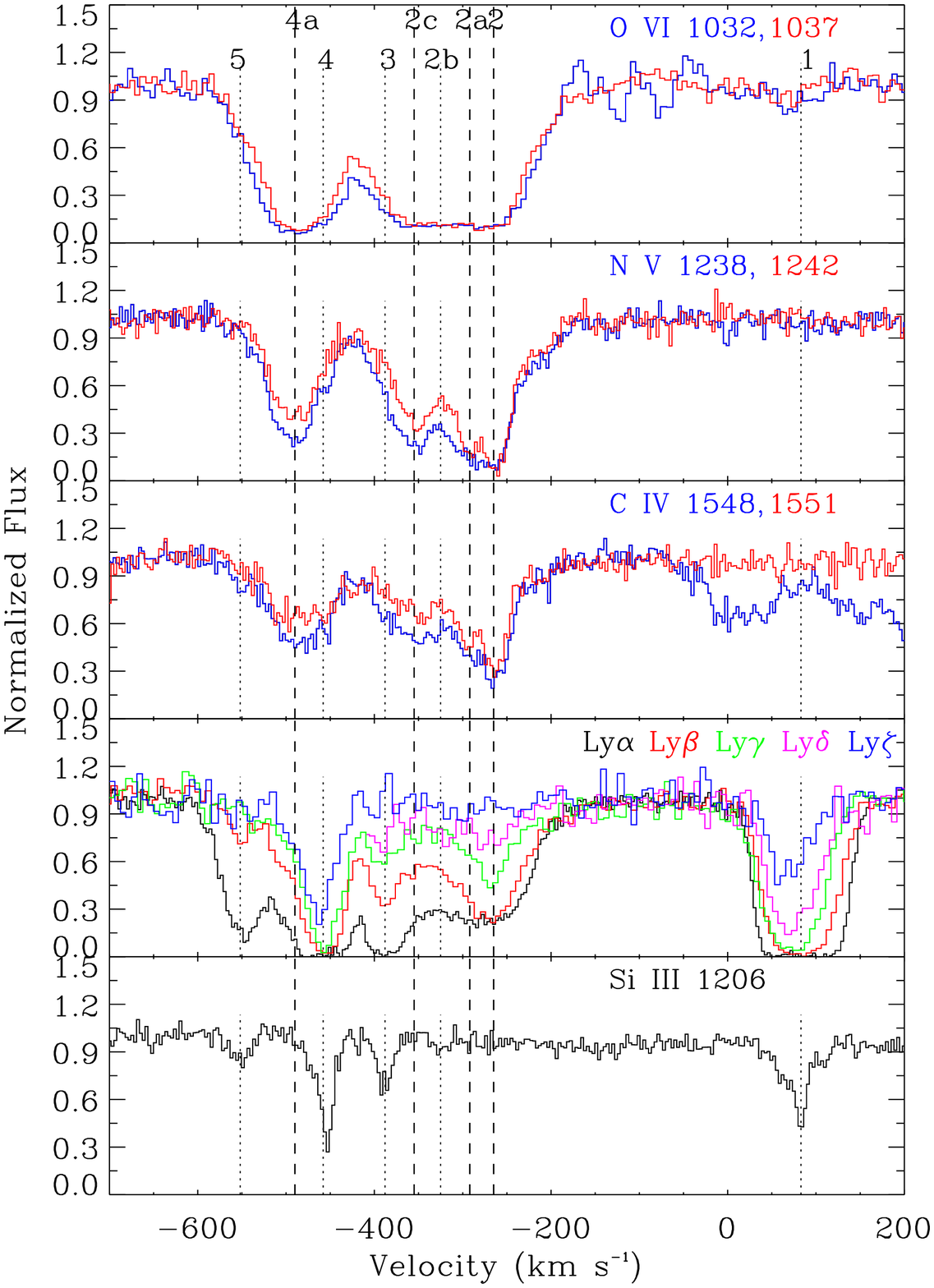}
\vspace*{0.35 in}
\caption{Normalized absorption profiles from 2003 May STIS and {\it FUSE} 
spectra.  The spectra are plotted as a function of radial velocity
with respect to the systemic redshift of the host galaxy.  The centroid
velocities of kinematic components associated with the AGN outflow are 
identified with dashed vertical lines.  Components identifying low-ionization
absorbers likely unrelated to the outflow (see text) are shown with dotted lines.
The difference in kinematic structure is evident in comparing the high-ionization
CNO doublets with Si\,{\sc iii}.  The Ly$\delta$ profile is not plotted at 
$v < -$400 km s$^{-1}$ due to contamination with Galactic absorption at these
velocities. \label{fig2}}
\end{figure}

      We adopt the component numbering system from SK04, which was based on the kinematic 
structure in Ly$\beta$.
In Figure 2, dotted vertical lines mark the centroids of the components in SK04 that
exhibit narrow absorption structure in low-ionization species in the 
current spectrum but which have no corresponding structure in the high-ionization 
CNO doublets.
Centroids of the components seen in the high-ionization lines are identified with
dashed lines; we have added component 2c to the SK04 system based on structure in the 
C\,{\sc iv} and N\,{\sc v} profiles. 
Measured centroid radial velocities and widths of the components are listed in Table 1.
The differences in ionization and kinematic structure between these two groups of components
strongly suggests they are physically distinct.
Based on their ionization, narrow widths, distinct centroid velocities,
and (in component 4) low density implied by the stringent upper limit 
on the C\,{\sc ii} column density in the excited fine-structure level, SK04 concluded
at least some of the low-ionization components are not associated with the AGN outflow.
Instead, they posited they are associated with gas located at relatively large
distances from the nucleus - perhaps from an interaction 
with the companion galaxy MCG$+$12-13-024, high-velocity clouds associated with the 
host galaxy of Mrk 279, or, in the case of component 1, in the interstellar medium 
of the host galaxy.

    In this study, we restrict our attention to the {\it bona fide} intrinsic
absorption, i.e., that presumed to be directly associated with an outflow from the AGN.  
We take this to include all absorption from the broad O\,{\sc vi}, N\,{\sc v}, and C\,{\sc iv} 
doublets; Figure 2 shows that any absorption associated with the
narrow low-ionization components will at 
most only effect the outer wings of the outflow components in these lines.
Conversely, Figure 2 shows Lyman line absorption from the low-ionization components
is strong and heavily blended with the intrinsic absorption components; thus,
we limit our analysis of H\,{\sc i} to the uncontaminated region, 
$v \approx -$300 to $-$200 km s$^{-1}$.

\section{The Doublet Method: Overview and Limitations}

  We present here a brief review of the standard technique for measuring 
UV absorption in AGN outflows and describe some limitations of this 
method to highlight the motivation for our method of analysis.
In earlier studies of intrinsic absorption, the red members of doublet pairs were
often found to be deeper than expected relative to the blue lines, based on their
intrinsic 2:1 optical depth ratios.
In many cases this was interpreted as due to partial coverage of the background 
nuclear emission by the absorbing gas, e.g., \citet{wamp93}, \citet{barl97}, 
\citet{hama97};
\citep[other possibilities are scattering from an 
extended region and emission from an extended source unrelated to the central engine 
of the AGN;][]{cohe95,good95,krae01}.
If partial coverage is not accounted for, the absorption ionic column densities
can be severely underestimated, which will dramatically affect the interpretation 
of the outflow.  The expression for the observed absorption that
includes the effects of line-of-sight covering factor ($C$) and optical
depth ($\tau$) is:
\begin{equation}
I(v) = (1 - C(v)) + C(v) e^{-\tau(v)},
\end{equation}
where $I$ is the normalized flux, and all quantities are written as a function of 
radial velocity, $v$.
Since the optical depths of the UV doublet pairs are in the simple 2:1 ratio,
equation 1 can be solved for the covering factor and optical depths 
of each doublet \citep{barl97,hama97}.
The resulting expressions, which we will refer to as the doublet solution, are:
\begin{equation}
C = \frac{I_r^2 - 2 I_r + 1}{I_b - 2 I_r + 1},
\end{equation}
\begin{equation}
\tau_r = -\ln(\frac{I_r - I_b}{1 - I_r}), 
\end{equation}
where $r$ and $b$ subscripts identify the red and blue members of the doublet, and the
equation for $\tau$ was derived by \citet{arav02}.  
These expressions can be evaluated for unblended (i.e., sufficiently narrow) 
absorption doublets with members that are individually resolved, and derived as
a function of radial velocity.
While this has provided a revolutionary advance in the study of
intrinsic AGN absorption, there are some key limitations to 
this method, described below.

\subsection{Multi-Component Nature of the Background Emission}

    An implicit assumption in the doublet solution is that 
the absorption is imprinted on a uniform, homogeneous background emission source,
since it allows for the solution of only a single $C$ and $\tau$.
However, the AGN emission is comprised of multiple, physically distinct 
sources, i.e., a continuum source and emission line regions (including
multiple kinematic components), which have different sizes, morphologies, and flux 
distributions.  Thus, in cases where the absorber only partially occults the total 
background emission, the distinct sources would be expected to have 
different line-of-sight covering factors, in general.
This possibility was first explored by Ganguly et al. (1999) for the continuum source
and broad emission line region (BLR), and was demonstrated in the
intrinsic absorption systems by Ganguly et al., \citet{gabe03}, \citet{hall03},
and SK04.

     To account for multiple discrete background emission sources, equation 1
can be expanded to give the normalized flux of the j$^{th}$ line
\begin{equation}
I_j = \Sigma_i  [ R_j^i  (C_j^i  e^{-\tau_j} + 1 - C_j^i) ], 
\end{equation}
where the i$^{th}$ individual emission source contributes a fraction 
$R_j^i = F_j^i / \Sigma_i [F_j^i]$ to the total intrinsic flux and has
covering factor $C_j^i$.
The {\it effective} covering factor for each line is the weighted combination
of individual covering factors, and can be written:
\begin{equation}
C_{j} = \Sigma_i C_j^i R_j^i
\end{equation}
These are expansions of the expressions given in \citet{gang99} 
to include an arbitrary number of emission sources.

     The multi-component nature of the background emission 
has several important implications for the analysis of AGN outflows:

$\bullet$  From equation 5 it can be seen that lines of the same ion 
could have different effective covering factors, which may introduce 
an error into the doublet equation.
This happens when the underlying emission fluxes differ from the spectral 
position of one line to the other and is illustrated in the N\,{\sc v} 
doublet absorption shown in Figure 3.
Here, the emission line flux underlying the blue member 
is $\approx$ 15\% greater than under the red line, while
the continuum flux under the two lines is identical.
The magnitude of this error depends on the slopes of the flux distributions
between the doublet lines and differences in individual covering factors
of the distinct emission sources.
\citet{gang99} showed this effect is typically small when considering
the continuum/BLR distinction, due to the gradual slope of the BLR; however,
if the doublet lines are near saturation, it could have a very
large effect since small flux differences correspond to large optical depth 
differences in these cases.  
Also, an underlying narrow emission line component could have a pronounced
effect on the solution \citep{arav02,krae02,gabe04}.

\begin{figure}
\includegraphics[width=8.5cm]{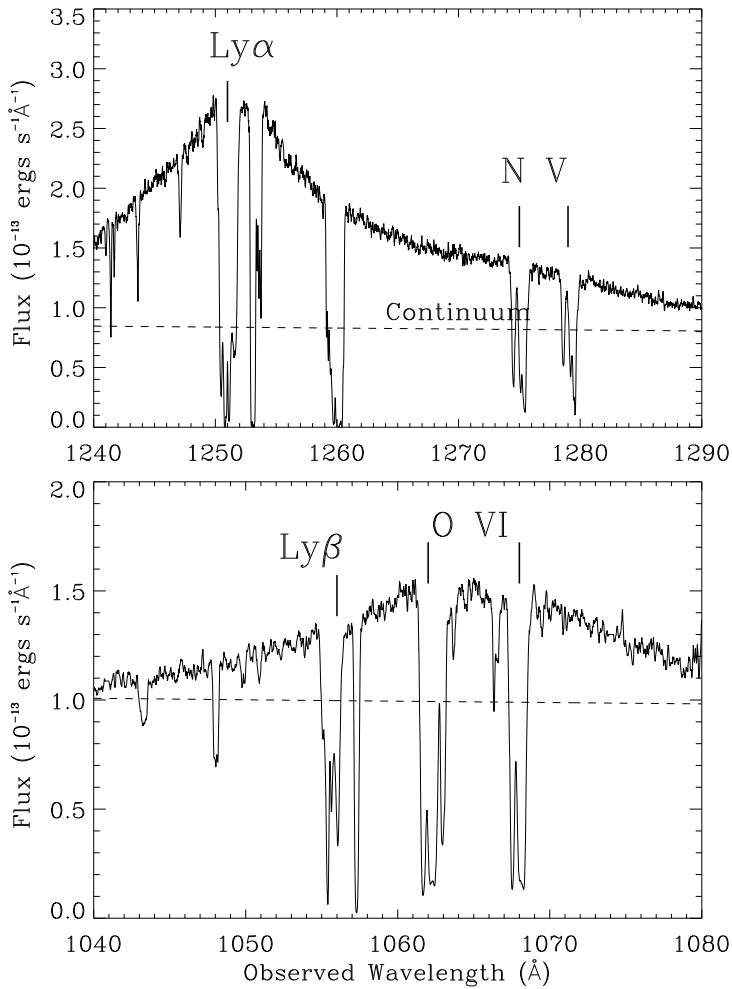}
\vspace*{0.35 in}
\caption{Spectrum of Ly$\alpha$ - N\,{\sc v} (top panel) and Ly$\beta$ - O\,{\sc vi} (bottom panel),
illustrating complexities in treating covering factors in intrinsic absorption measurements.
The continuum flux level, plotted as dashed lines, is essentially identical for each pair of lines, 
while the emission-line fluxes underlying each line differs greatly.
This has important implications for the relative {\it effective} covering factors.
Additionally, the nature of the BLR emission underlying these lines is complex: Ly$\beta$ 
absorption lies on the high-velocity blue wing of O\,{\sc vi}, while N\,{\sc v} absorption 
has a contribution from the red wing of the Ly$\alpha$ BLR.\label{fig3}}
\end{figure}

$\bullet$  Without separation of the covering factors of the distinct sources,
covering factors derived from a doublet pair cannot be applied to 
measure column densities of other lines.  This is evident in Figure 3; 
clearly the effective covering factor derived for N\,{\sc v}
is not applicable to Ly$\alpha$ if the individual continuum and emission-line 
covering factors differ since Ly$\alpha$ has much more underlying line flux.
Similarly, Ly$\beta$, plotted in the bottom panel in Figure 3, will not generally 
have the same effective covering factor as Ly$\alpha$, due to the different emission line 
contributions under each line.
However, if the individual covering factors are known,
effective covering factors can be constructed for any line using equation 5.
This is important for measuring singlet lines or contaminated multiplets that
have no independent measure of the covering factor.

$\bullet$  Finally, the doublet solution misses potentially important, unique 
constraints on the absorption and emission geometry.
For example, combined with estimates of the
sizes of the individual sources derived from other techniques, the individual 
covering factors constrain the size of the absorber, e.g. \citet{gabe03}, and the 
relative location of the different emission components and absorber as projected
on the plane of the sky.
The individual covering factors can also serve as a unique probe of more detailed 
geometry of the background emission.
Consider for example the O\,{\sc vi} -- Ly$\beta$ spectrum shown in Figure 3.
The O\,{\sc vi} doublet absorbs its own emission line flux at the blueshifted
velocity of the outflow ($v_{BLR} \approx -$600 to $-$ 200 km~s$^{-1}$), 
while the Ly$\beta$ absorber sits primarily on the high-velocity
blue wing of the O\,{\sc vi} BLR at $v_{BLR} \approx -$2200 km~s$^{-1}$.
The Ly$\beta$ line emission is relatively weak.
Similarly, there is a contribution from the extreme red wing of the Ly$\alpha$ BLR
profile under the N\,{\sc v} absorption.
Constraints on the emission line covering factors for these lines could
be used to probe the kinematic-geometric structure of the BLR; the
absorber can thus serve as a filter to view and explore the background AGN sources.

\subsection{Systematic Errors in the $C$ -- $\tau$ Solutions}

   Another limitation is that the doublet method always gives a solution, 
but it is often difficult to gauge its accuracy due to the
non-linear dependency of the solution on measurement errors.
To explore this, we have generated synthetic absorption profiles that include
random fluctuations simulating spectral noise, and calculated $C$ and $\tau$ using 
the doublet equations.  Illustrative results are shown in an Appendix, 
whereas a complete quantitative treatment will be presented in a later paper.
We find there are systematic errors in the solutions that can give misleading
results.  These errors are not random about the actual value, but rather systematically
underestimate the actual covering factor, as seen in the Appendix; the discrepancy in the
solution increases with weaker absorption doublets and decreased S/N.
Additionally, \citet{gang99} demonstrated that the finite instrumental
line spread function can lead to further systematic errors in the doublet solution.

\section{Optimization Fitting of the Intrinsic Absorption:  
Lyman Series and CNO Doublet Global Line Fits}

  Motivated by the limitations of the traditional doublet method described 
above, we introduce here a different approach for measuring intrinsic
absorption.  The underlying principle is to increase the number of 
lines that are simultaneously fit in order to (a) explore additional
parameters contributing to the formation of observed absorption troughs and (b)
overconstrain the set of equations.
This allows for the treatment of more complex, physically
realistic scenarios of the absorption-emission geometry.
By minimizing errors to simultaneous fits of multiple lines, 
noise in the spectrum will generally be smoothed out, 
in contrast to the erratic behavior of the doublet solution demonstrated
in \S 3.2 and the Appendix.

\subsection{Formalism}

   Our fitting algorithm employs the Levenberg-Marquardt non-linear 
least-squares minimization technique to solve equation 4 for specified 
absorption parameters ($C_j^i$, $\tau_j$)\footnote{Using software provided by C. Markwardt, 
http://cow.phys.wisc.edu/$\sim$craigm/idl/idl.html, which is based on the MINPACK-1 
optimization software of J. Mor\`e available at www.netlib.org}. 
It is similar in principle to that used in SK04 to analyze the Lyman lines 
in earlier spectra of Mrk 279.  Given a total of n 
observed absorption lines ($I_j$) as constraints, up to n$-$1 parameters can be modeled.
No a priori assumptions are made about the kinematic 
distribution of the covering factors and optical depths of the absorbing 
material (e.g., Gaussian) -- indeed, one goal is to solve
for the velocity-dependent absorption parameters to
constrain the kinematic-geometric structure of the mass outflow.  
Thus, we derive fits to the absorption equations for each
velocity bin.  This also avoids errors in the solutions resulting from 
averaging over variable profiles.  
The algorithm minimizes the $\chi^2$ function, with each data 
point appropriately weighted by the 1 $\sigma$ errors, which are 
a combination of spectral noise and estimated uncertainties in 
fitting the intrinsic underlying fluxes.  For the latter, the 
continuum flux uncertainties were determined from the residuals 
between the power-law model and the line-free regions of the spectrum.
We estimated uncertainties in the emission-line fluxes by testing 
different empirical fits over the absorption features, finding the range
that gave what we deemed reasonable line profile shapes.

    The key requirement in employing this technique is to link multiple absorption 
lines for simultaneous fitting.  There are two general ways to do this:

$\bullet$  {\it Lines from the Same Energy Level:}  The most straightforward way is to fit 
all available lines arising from the same ionic energy level, thereby eliminating 
uncertainties in ionic abundances or level populations.
If the relative underlying fluxes from distinct emission sources
differs between the lines, the individual covering factors of those emission sources
can be derived.
In the subsequent analysis, we fit all of the uncontaminated Lyman series 
lines in Mrk 279.  These lines are ideal because they
span a very large range in optical depth and have significantly 
different amounts of underlying emission-line flux \citep[see][]{gabe03}.
Additionally, the full set of lines is accessible in low redshift AGNs with 
combined {\it FUSE} and STIS spectra.
We note the Fe\,{\sc ii} UV multiplets, which appear in a small fraction of 
AGN absorbers \citep{deko01,krae01}, are another 
promising set of lines for this analysis.

$\bullet$  {\it Global Fitting Approach:}  The second approach involves linking lines
from different ions (or any group of lines arising from different levels) by placing physically 
motivated constraints on their absorption parameters.
In our analysis of Mrk 279 below, we fit the six combined lines of the 
O\,{\sc vi}, N\,{\sc v}, and C\,{\sc iv} doublets by assuming they
share the same covering factors.
Another potential application of this method is to link the absorption for a given line 
in spectra from different epochs via assumptions about the relative values of 
the absorption parameters between epochs (e.g., assuming the covering factors did not 
change).  The validity of the assumptions used to link the equations  
can then be tested by the result of the fit.

\subsection{Covering Factor and Optical Depth Solutions for Mrk 279}

    For the intrinsic absorption in Mrk 279, 
we independently fit the two groups of lines described above:  the 
Lyman series lines and the combined CNO doublets, i.e. global fit.
For each set of lines, we tested two different models of the absorption covering factor.  
In model A, a single covering factor was assumed to describe all lines, i.e., no distinction 
was made between the different emission sources.  
In model B, independent covering factors for the continuum source and emission lines, 
$C^c$ and $C^l$, were assumed.
In this case, the general expressions in equations 4 and 5 reduce to those in 
\citet{gang99}.

    For the Lyman lines, the solvable range is limited 
to $-$300 $\lesssim v \lesssim -$200~km~s$^{-1}$, due to blending with
the narrow, low-ionization components in the high velocity 
region of the outflow (see \S 2.2).
This is due to the failure of equation 4 where multiple absorption components with
different covering factors contribute in the same velocity bin; there 
is no straightforward way to disentangle how the different absorbers 
overlap as projected against the background emission sources.
Figure 2 shows Ly$\alpha$, Ly$\beta$, and Ly$\gamma$ give the best constraints
for the Lyman series analysis, exhibiting clean well-defined absorption profiles
at relatively high S/N.
Weak absorption in Ly$\delta$ is also present in component 2, while 
Ly$\epsilon$ is contaminated with Galactic H$_2$ absorption 
and thus omitted from the fitting.
Ly$\zeta$ is not detected in components 2 -- 2a within the limits
of the spectral noise, thus to reduce the effect of noise on the solution,
we set the normalized flux to unity for this line.
All lines of higher order than Ly$\zeta$ were omitted from the analysis since
they exhibit no intrinsic absorption and provide no additional constraints.
Thus, there are five lines as constraints to fit the two and three free 
parameters for the Lyman solution in model A and B, respectively.

    For the global fitting of the O\,{\sc vi}, N\,{\sc v}, and C\,{\sc iv} doublets,
the absorption equations were linked by assuming their covering factors are
equal (separately for the continuum and emission lines in model B).
The optical depth of each ion is a free parameter.
As described in \S 2.2 and seen in Figure 2, these lines are not strongly contaminated 
by the narrow, low-ionization absorption that dominates the Lyman lines; at most, there is
only weak contamination in the wings of the broad, intrinsic absorption 
features by these low-ionization systems.

\begin{figure}
\includegraphics[width=8.5cm]{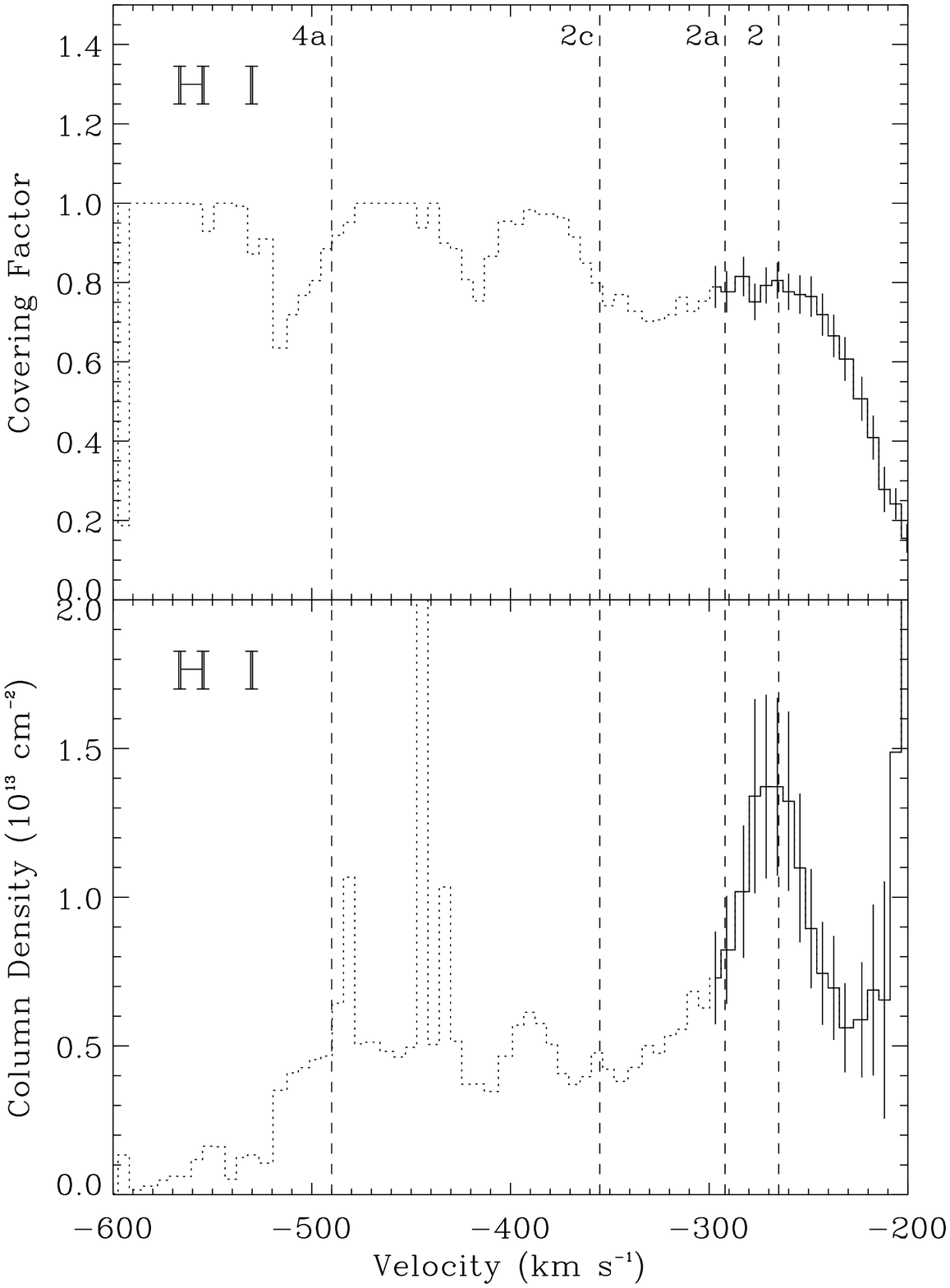}
\vspace*{0.35 in}
\caption{Best-fit covering factor and column density/optical depth solutions from $\chi^2$ 
minimization for the single covering factor geometric model (model A). 
Top panels show $C$ solutions, solved independently
for the Lyman series (left panels) and combined CNO doublets (right). 
Bottom panels show the H\,{\sc i} column density and C\,{\sc iv} (dashed histogram), 
N\,{\sc v} (dotted), and O\,{\sc vi} (solid) optical depth solutions.
The contaminated, and thus unreliable, region of H\,{\sc i} absorption is plotted
with dotted histograms. \label{fig4a}}
\end{figure}  

\addtocounter{figure}{-1}

\begin{figure}
\includegraphics[width=8.5cm]{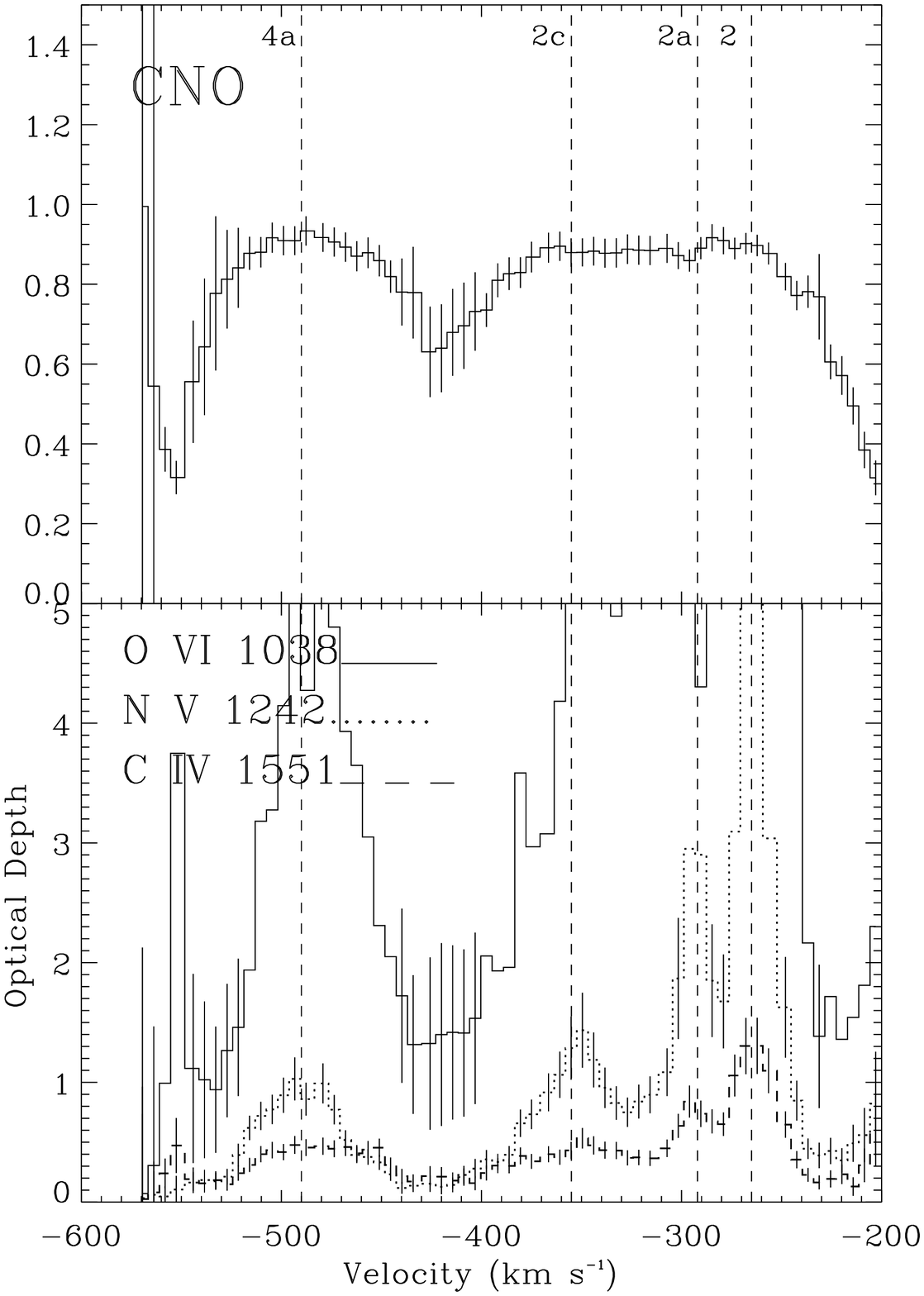}
\vspace*{0.35 in}
\caption{ \label{fig4b}}
\end{figure}  

\begin{figure}
\includegraphics[width=8.5cm]{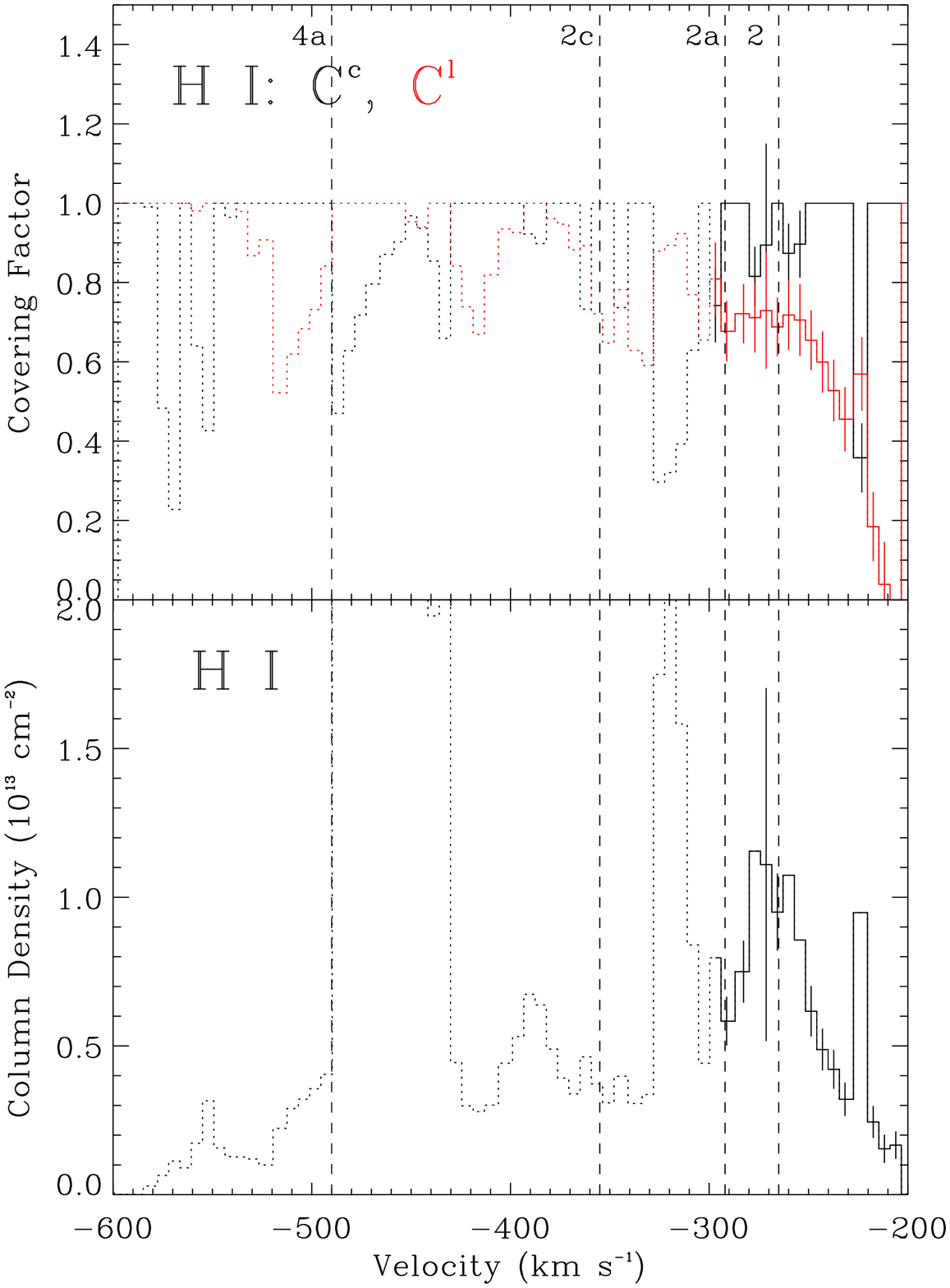}
\vspace*{0.35 in}
\caption{Same as Figure 4 for the two covering factor geometric model (model B).
Continuum covering factors are plotted in black and emission-line covering factors
in red.  \label{fig5a}}
\end{figure}  

\addtocounter{figure}{-1}

\begin{figure}
\includegraphics[width=8.5cm]{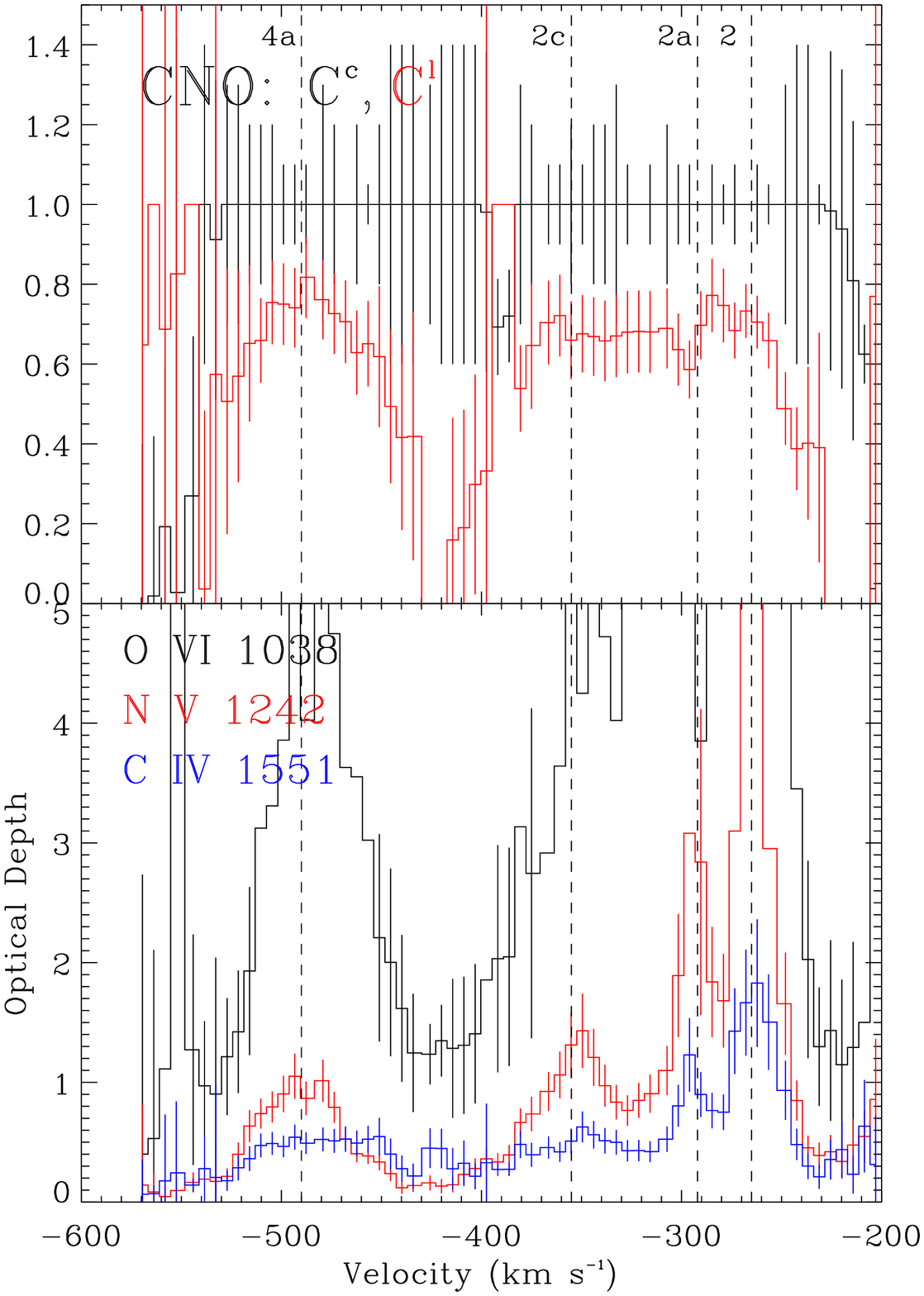}
\vspace*{0.35 in}
\caption{ \label{fig5b}}
\end{figure}

\begin{figure}
\includegraphics[width=8.5cm]{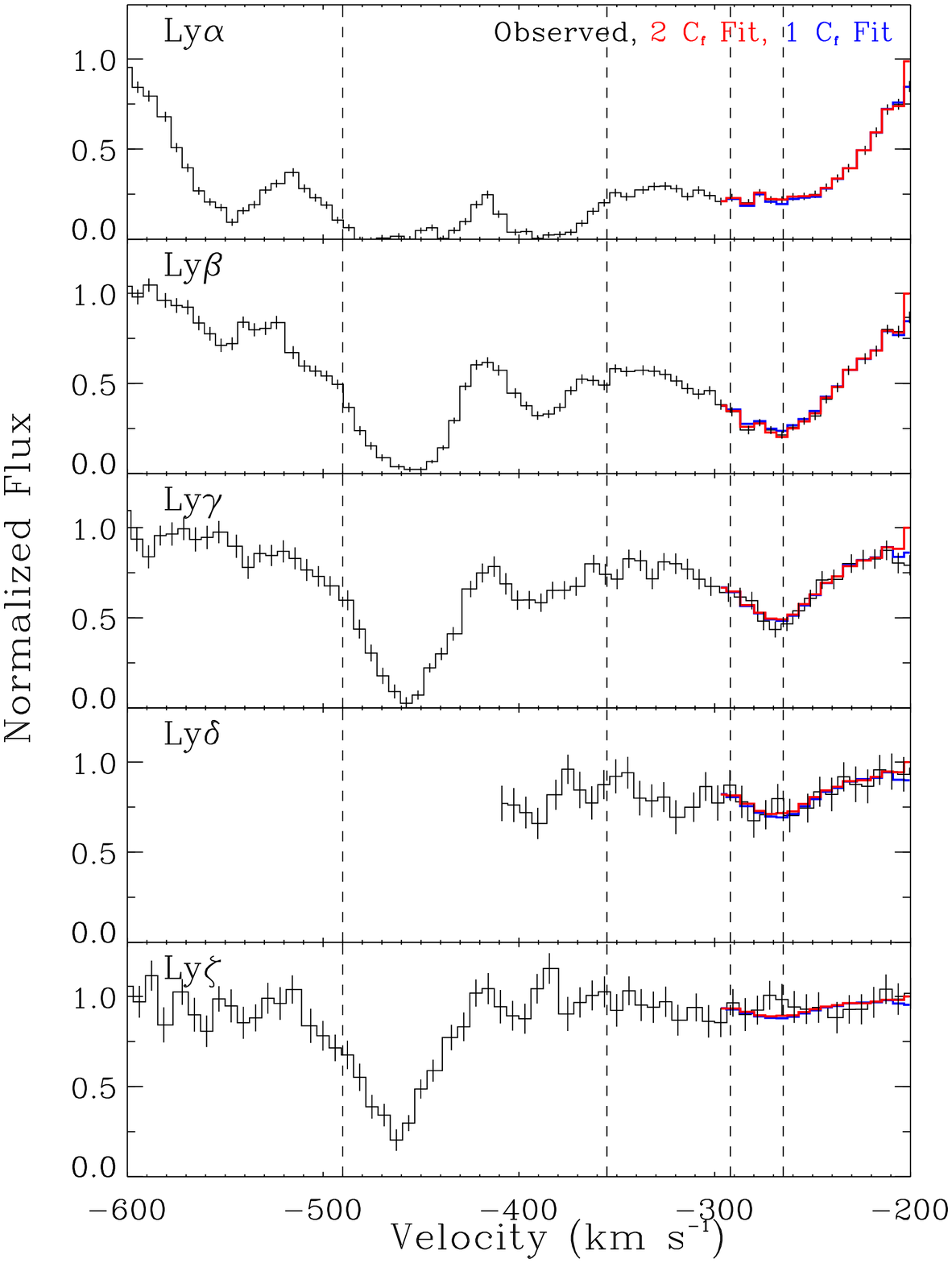}
\vspace*{0.35 in}
\caption{Best-fit intrinsic absorption profiles for the Lyman lines (left) and CNO
doublets (right) from the two geometric models.  The normalized observed 
spectrum is plotted in black, model A profiles are in blue, and model B profiles in red.
The model profiles were derived from the best-fit covering factors and column densities
shown in Figures 3 and 4 using equation 4.
The contaminated region of the Lyman line fit (dotted lines
in Figures 3,4) is not plotted. \label{fig6a}}
\end{figure}  

\addtocounter{figure}{-1}

\begin{figure}
\includegraphics[width=8.5cm]{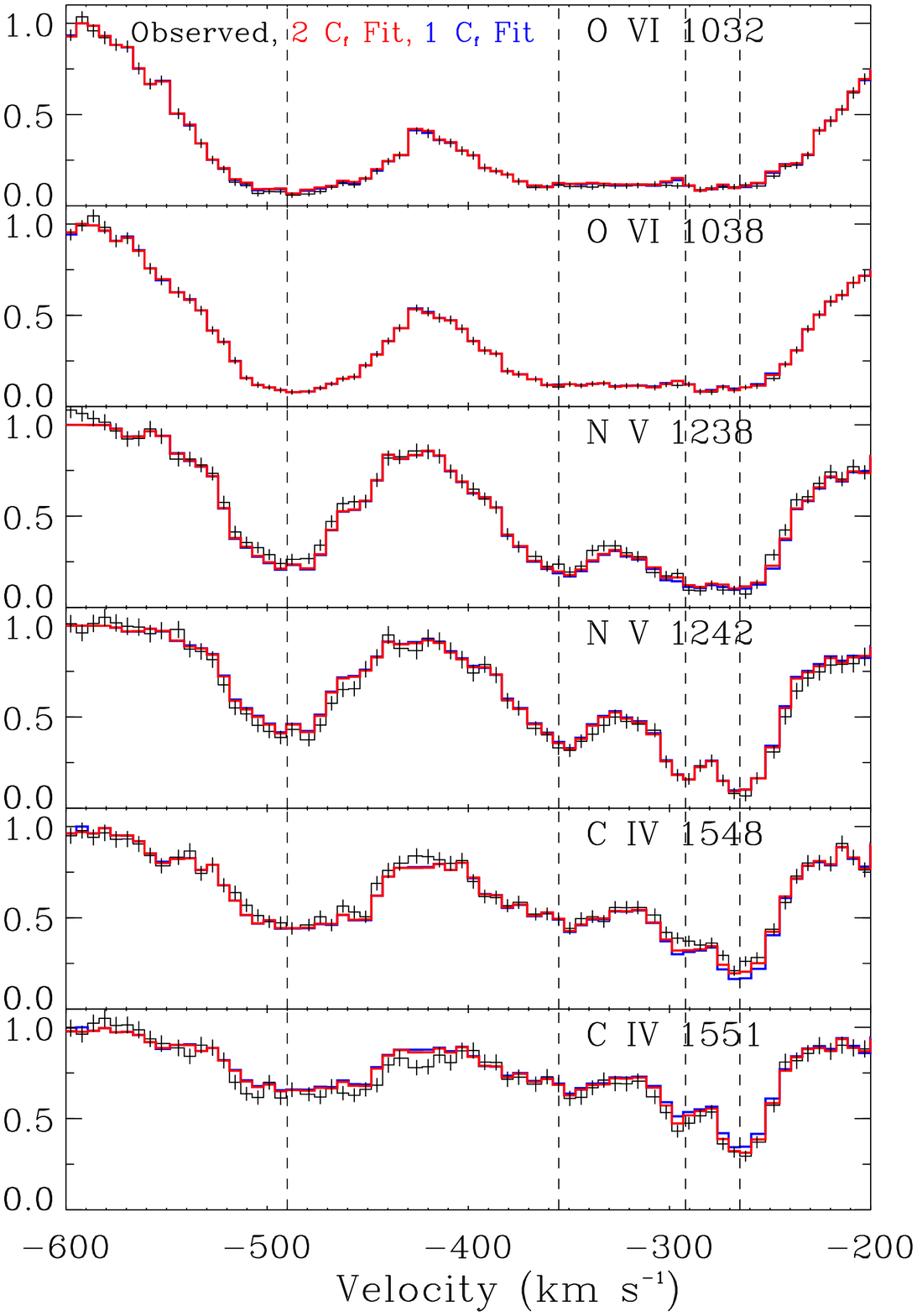}
\vspace*{0.35 in}
\caption{ \label{fig6b}}
\end{figure}

   Best-fits to the covering factors and optical depths (or equivalently 
column densities) for both line groups are shown in Figure 4 (model A) and Figure 5 (model B).  
In all cases, the parameter space search for the covering factors was
restricted to the physically meaningful range, 0~$\leq C^i \leq$~1.
The plotted error bars represent the formal 1 $\sigma$ statistical errors in
the best-fit parameters.
Specifically, these correspond to values giving $\Delta \chi^2 =$1, 
and were computed from the diagonal elements of the 
covariance matrix for the optimal fit \citep{bevi69}.
For cases where the covering factor solution is a boundary 
value (0,1), e.g. much of the $C^c$ solution for the CNO doublets in
model B, no covariance matrix elements are computed 
for that parameter since it is not a minimum in the solution.
For these cases, we estimated uncertainties by 
deriving solutions for models keeping the parameter fixed, and finding the 
value giving $\Delta \chi^2 =$1 from the best-fit
solution at the boundary value.
To ensure the computations did not erroneously stop at local minima,
we generated solutions using different starting points for the parameter
search space; identical results were found in all cases.
The fitted profiles, derived by inserting the best-fit solutions into equation 
4, for both models (model A in blue, model B in red) are compared with the 
observed normalized absorption profiles in Figure 6.

\subsection{Errors and Uncertainties}

    There are some possible errors in this fitting that should be mentioned.
First, the emission lines are treated as all arising from a single component. 
Line emission from distinct kinematic components, i.e. broad, intermediate, 
narrow line regions, that are covered by different fractions by the absorbers
could introduce an error into the solution (although, the narrow line region in
Mrk 279 is relatively weak).  
Also, there are cases where absorption features sit on the BLR emission from
different lines and thus sample different velocities of the BLR profile (i.e., 
Ly$\beta$ and N\,{\sc v} described in \S 3.1 and shown in Figure 3).
This could introduce an error into the solution if there are spatial
inhomogeneities in the BLR gas as a function of velocity.
Second, the area on the sky sampled by the {\it FUSE} aperture is four orders 
of magnitude greater than the STIS aperture.  While this has no consequences 
for the continuum and BLR emission, which are unresolved and much smaller than the 
0.$\arcsec$2 STIS aperture, any extended emission might effect the solutions.
This could include scattering of nuclear emission by an extended scatterer 
\citep[e.g.][]{krae01} or extended O\,{\sc vi} NLR emission.
Finally, we have assumed the absorption optical depth of each line ($\tau_j$) is uniform 
across the lateral extent of the emission sources, and thus the absorber is completely 
homogeneous \citep[see][]{deko02}.  Models departing from this assumption 
are presented in \citet{arav04}.

\section{Discussion and Interpretation}

\subsection{Favored Absorption Model: Independent Continuum and 
Emission Line Covering Factors}

  Figure 6 shows both geometric models are able to match 
the intrinsic absorption features well at most outflow velocities, 
indicating the solutions for these models and 
parameters are degenerate over much of the profiles.
There are some regions fit somewhat better by
model B, particularly C\,{\sc iv} in the low-velocity outflow component.
However, there is additional, stronger evidence that supports the two covering
factor model for the outflow, as outlined below.

$\bullet$ {\it Consistent Covering Factor Solutions From the Independent
Lyman Line and Global CNO Doublet Fitting Methods:}  
Figure 5 shows that both the Lyman and global CNO doublet best-fit
solutions are consistent with full coverage of the continuum source over most 
of the velocity range with a valid Lyman solution.
Further, the independent solutions for the two groups of lines are 
plotted together in Figure 7 for a direct comparison; the emission line
covering factor profiles computed in model B are shown in the top 
panel and the single covering factors derived in model A are in 
the bottom panel.  The H\,{\sc i} and CNO $C^l$ fits are nearly
identical over most of the core region of the absorption, 
with $C^l \approx$ 0.7 (although in the wing of the absorption, 
$v > -$250 km~s$^{-1}$, the solutions diverge somewhat).  In contrast, 
the Lyman method solution in model A is systematically less than the global doublet 
solution by about 0.1 -- 0.15 at all velocities.

\begin{figure}
\includegraphics[width=8.5cm]{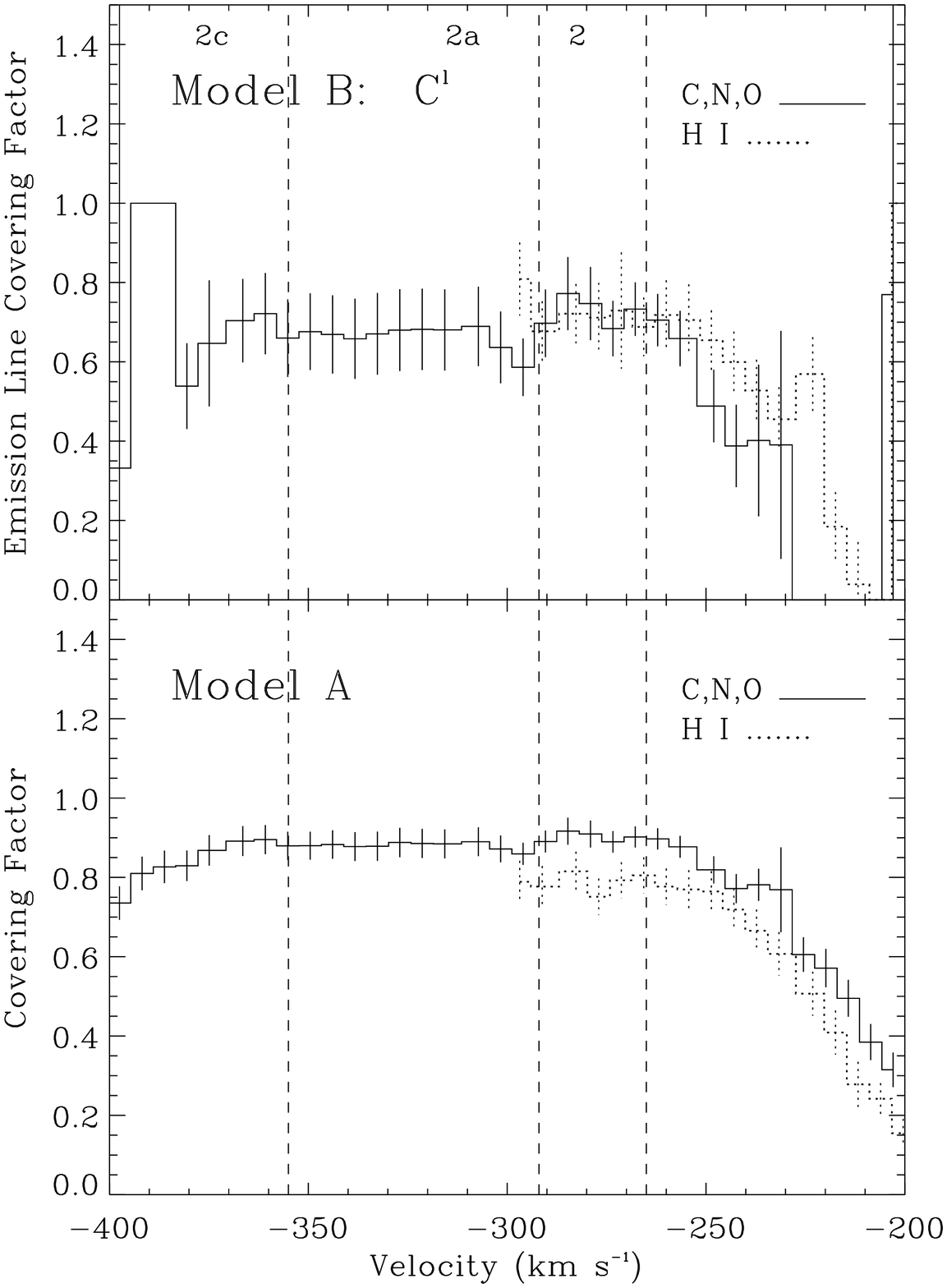}
\vspace*{0.35 in}
\caption{Comparison of covering factor solutions for the
two independent methods -- Lyman and global CNO doublet lines.
Best-fit solutions to the emission-line covering factors
from model B are shown in the top panel. The single covering factor
solutions from model A are in the bottom panel. \label{fig7}}
\end{figure}

$\bullet$ {\it Consistency with the Emission Source Sizes:}  
The continuum and emission line covering factor solutions in model B
are also physically consistent with our understanding of the
sizes of those emission sources.
Reverberation mapping studies of AGNs have shown the BLR is substantially
larger than the ionizing continuum source.  
These studies measure the BLR to be several to tens of light-days across 
\citep[e.g.][]{pete98,kasp00},
while the UV continuum source is likely at least an order of magnitude
smaller in size and possibly much smaller \citep[e.g.][]{laor89,prog03}.
Thus, the solution of a fully covered continuum source and partial
coverage of the emission lines is consistent with the nuclear emission geometry.
The fact that this result was arrived at separately with the independent
Lyman and global CNO fits is compelling.

$\bullet$ {\it Variability in Ly$\alpha$ Absorption:}  In Figure 8, 
the normalized Ly$\alpha$ profile 
from our observation is compared with the STIS spectrum
obtained a year earlier, showing the absorption was shallower in 
the previous epoch by $\sim$ 0.15 in normalized flux units.
This variability is not due to a lower H I column density; the
strength of Ly$\beta$ in a contemporaneous {\it FUSE} observation
indicates Ly$\alpha$ was saturated in the earlier epoch (SK04), as it
is in the 2003 spectrum, hence the profiles in both epochs simply
delineate the unocculted flux.  Thus, a change in covering factor must be 
responsible for the observed variability.
Noting that the emission line-to-continuum flux ratio was greater in the 2002
spectrum than in 2003, we tested if the difference
in covering factor of the respective sources could explain the variability.
To this end, we constructed a synthetic absorption profile for Ly$\alpha$
in the 2002 epoch based on the results from our
analysis of the 2003 spectrum.
Specifically, we set $C^c =$1 at all velocities, and solved
for $C^l$ in the 2003 Ly$\alpha$ profile.
The observed continuum and emission line fluxes in the 2002 spectrum were then
weighted by these covering factors according to equation 4.
The resulting model profile, shown in the bottom panel of Figure 8, 
matches the observed absorption very well. 
This provides a natural explanation for the variability in the 
Ly$\alpha$ absorption profile:
the covering factors of each source individually remain the same, but a change in
the {\it effective} covering factor occurs due to different relative
strengths of the distinct background emission components.

\begin{figure}
\includegraphics[width=8.5cm]{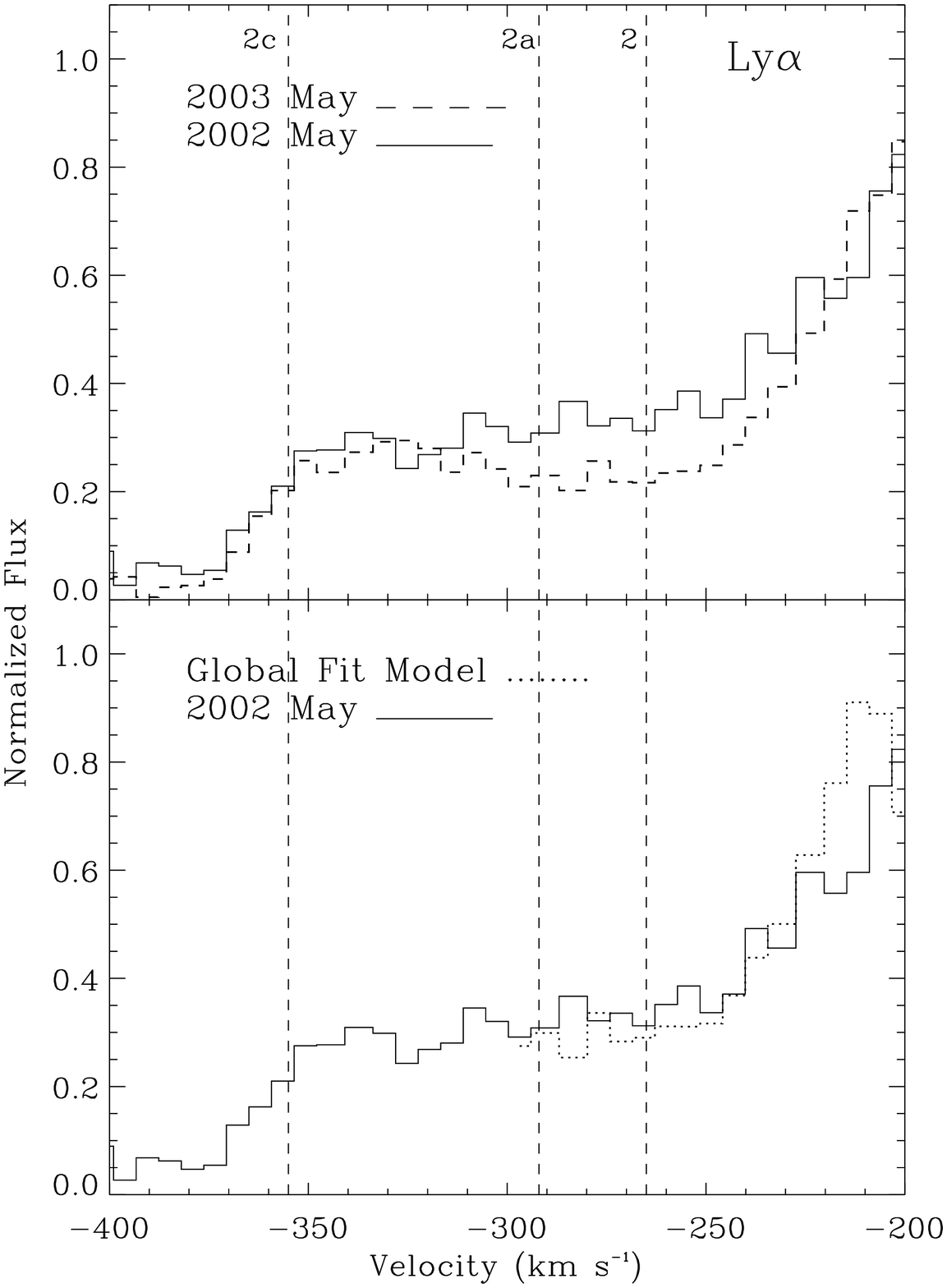}
\vspace*{0.35 in}
\caption{Variability in Ly$\alpha$ absorption.  The top panel shows Ly$\alpha$ 
absorption in components 2--2a was shallower in a May 2002 STIS spectrum (solid histogram) 
than in May 2003 (dashed).  The bottom panel shows a model of the 2002 profile (dotted), 
constructed with individual covering factors derived from the 2003 spectrum (Figure 5), 
matches the observed profile well.
This indicates the variation in absorption depth is likely due to 
a change in {\it effective} covering factor resulting from different
relative strengths of the continuum and emission-line fluxes 
in the two epochs.\label{fig8}}
\end{figure}

\subsection{Comparison of Solutions from Different Methods}

   The covering factor and optical depth solutions for C\,{\sc iv},
N\,{\sc v}, and O\,{\sc vi} from the two-$C$ global fit (model B) 
are compared with the single doublet solutions
(equations 2 and 3) in Figure 9.  The global fit effective covering factors are 
weighted combinations of the individual covering factors
shown in Figure 5, derived using equation 5.
In Figure 9, solid red lines show $C$ profiles 
for the short-wavelength members of each doublet and dotted 
lines show results for the long-wavelength lines; 
they are nearly identical in all cases, indicating negligible errors
in the doublet solution due to different contributions from underlying
line emission (see first bullet in \S 3.1).

\begin{figure}
\includegraphics[width=8.5cm]{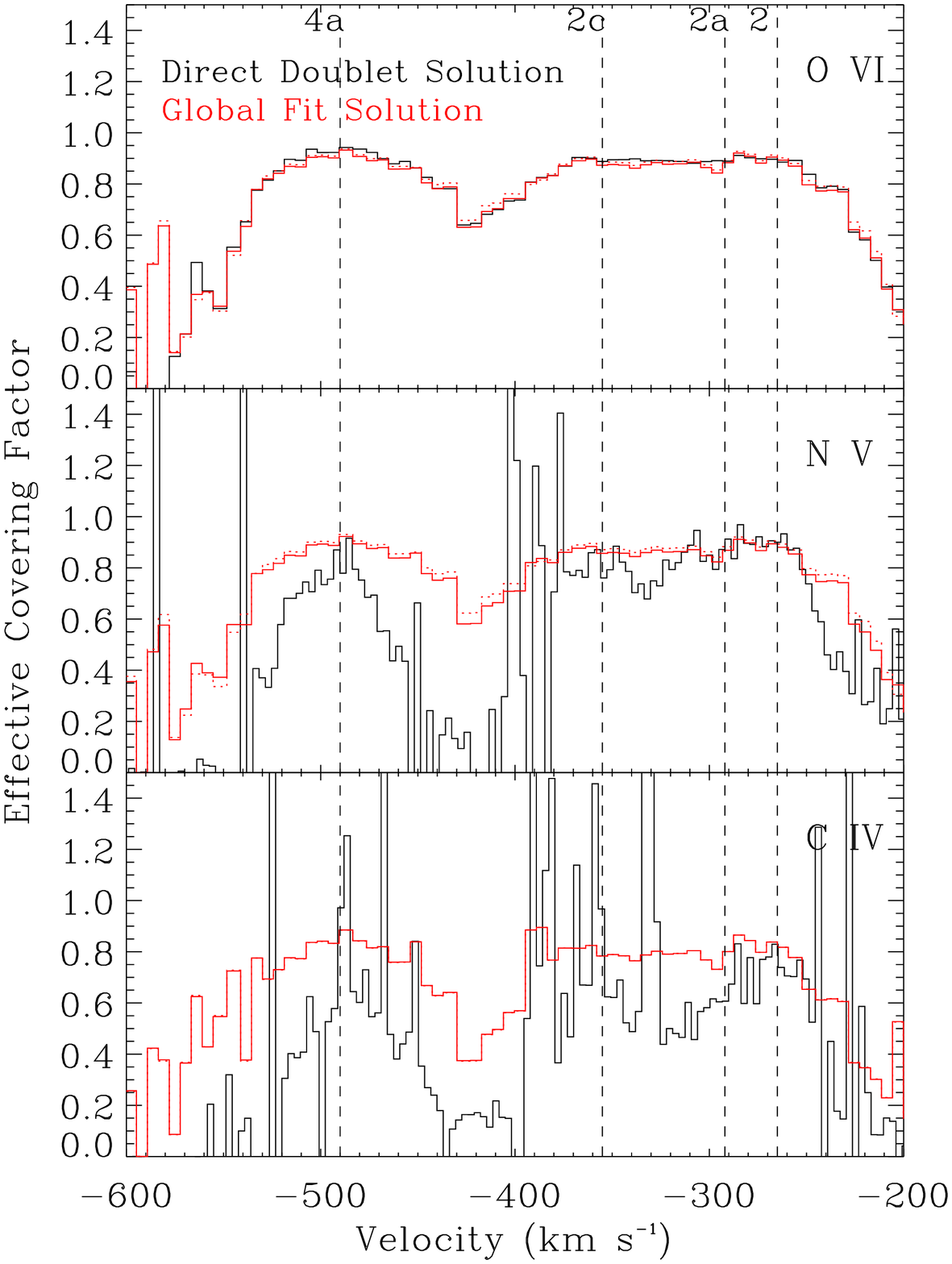}
\vspace*{0.35 in}
\caption{Comparison of covering factor and optical depth solutions 
derived with the global-fit (red) and single doublet (black) methods.
Left panels: Covering factors from the global-fitting method are the
effective values derived by weighting the individual covering
factors in model B by their respective fluxes. 
The short-wavelength doublet members are
plotted with solid red lines and the long-wavelength members with
dotted lines.  Right panels: Optical depths for the long-wavelength lines of the 
doublets. \label{fig9a}}
\end{figure}  

\addtocounter{figure}{-1}

\begin{figure}
\includegraphics[width=8.5cm]{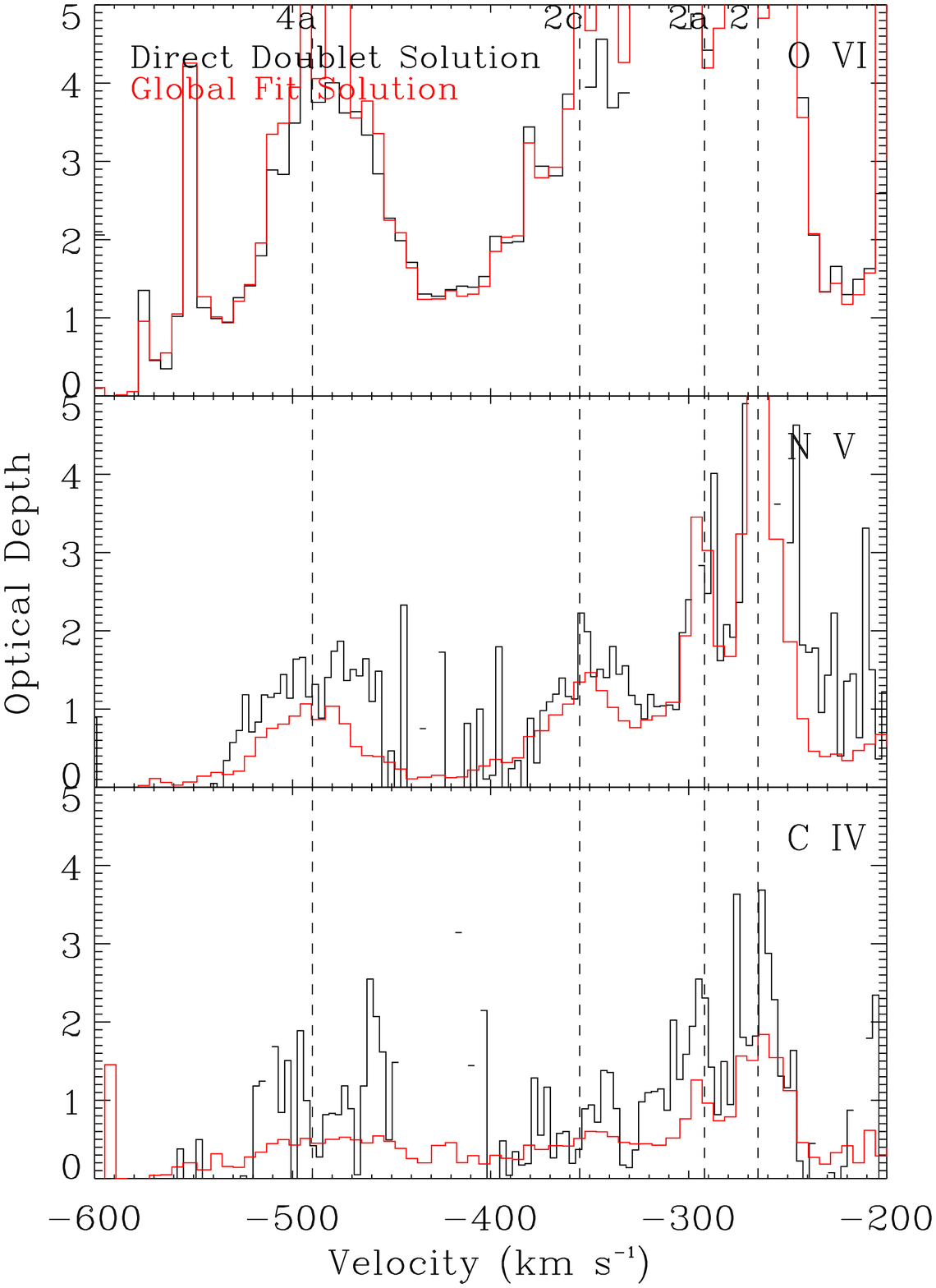}
\vspace*{0.35 in}
\caption{ \label{fig9b}}
\end{figure}

    Figure 9 shows the solutions from the two methods differ significantly in
some regions, with important implications for the kinematic-geometric and
-ionization structure in the outflow.
For example, the doublet covering factor solution, plotted in black, exhibits 
much more velocity-dependence than the global fit.
This is seen, for example, in the wings of C\,{\sc iv} and N\,{\sc v} in component 4a
and the red wing of component 2, where $C$ decreases strongly in the 
wings, while the optical depth profiles show little or no relation to the absorption 
trough structure when compared with Figure 2.  
Further, the kinematic structure in these ions is much 
different from the O\,{\sc vi} solution.
In these cases, the doublet solution implies the observed absorption 
profiles are determined almost entirely by velocity-dependent covering factor.
Additionally, in some regions, particularly in C\,{\sc iv}, the doublet solutions 
for $C$ and $\tau$ exhibit sharp fluctuations over small velocity 
intervals that do not coincide with structure in the observed absorption features,
nor with the solutions from the other ions (e.g., components 2a -- 2c, and 4a).  

    Comparison with the observed profiles in Figure 2 shows these effects 
in the doublet solution are strongly correlated with the absorption strength. 
Smaller derived covering factors are associated with weaker absorption, 
both in the wings of individual kinematic components and in overall
absorption features, such as C\,{\sc iv} components 2c and 2a. 
As discussed in the Appendix and \S 3.2, synthetic absorption 
profiles show the doublet solution suffers from systematic errors consistent 
with this general trend -- spectral noise causes an
underestimate of the covering factor.
The simulations we present in the Appendix indicate the solutions become unreliable 
for cases where $\tau \lesssim$ 1 in the blue 
doublet member, and increasingly so for lower $\tau$.
This, combined with the large, seemingly random fluctuations seen in certain 
regions suggests some features in the doublet solutions in Figure 9 are artifacts due
to noise and not real features in the outflow.
However, one puzzling aspect of the solutions, in comparison to the results in the
Appendix, is that there are not more negative values for the covering factors.
The simulations predict that for sufficiently low optical depth, there is a high
probability that $C <$0.
This should be especially apparent for intermediate values of $\tau$, such as the
middle two models presented in the Appendix, where dramatic fluctuations
between large positive and negative values are expected.
It is unclear why this is not more prevalent in the solutions in Figure 9.

   In Table 2, we compare integrated column densities measured from the different 
solutions:  the two geometric coverage models presented in \S 4.2, plus the 
traditional single doublet solution for the CNO doublets.
The integrated column densities for the doublet solution are larger than 
the global fits in all cases.
In component 2$+$2a, the doublet solution is 60\% greater than the 2-$C$ model for
C\,{\sc iv}, while in component 4a, it is 45\% and 70\% greater for C\,{\sc iv} and
N\,{\sc v}, respectively.
Comparing the two global fit models, the CNO column densities from model B
are greater than the single $C$ model by 0 -- 30\% , while
the H\,{\sc i} column density is 50\% less.

\subsection{Constraints on the Nuclear Absorption -- Emission Geometry}

    Here we explore constraints on the nuclear absorption and
emission geometry available from the covering factor solutions derived in \S 4.2.
Full analysis of the physical conditions in the outflow, using photoionization
modeling of the combined UV and X-ray absorption, will be presented in a
future study.

     As discussed in \S 5.1, we adopt the 2-$C$ global fit (model B) for the 
covering factors and column densities.
The covering factor solutions in Figure 5 are consistent with the UV continuum
source being fully in our sight-line through the absorber, for all kinematic
components, while the emission lines are only partially so.
This constrains the relative line-of-sight geometry of the absorbers and nuclear emission
sources and places a lower limit on the transverse size of the outflow.
Monitoring of time lags between BLR and continuum variations in Mrk 279 by 
\citet{maoz90} gives the size of the BLR, $R_{BLR} =$ 12 light days.
Thus, the projected size of the UV absorbers on the plane of the sky, against
the AGN emission, is at least
$(C^l)^{1/2} \times R_{BLR}$, or $\approx$10 light days in the cores of the
absorption components, and possibly decreasing in the wings.

\begin{deluxetable*}{lcccccccc}
\tablewidth{0pt}
\tablecaption{Integrated Ionic Column Densities\tablenotemark{a} \label{tbl-2}}
\tablehead{
\colhead{} & \multicolumn{2}{c}{Model A: 1 $C$ Fit}  &  \multicolumn{2}{c}{Model B: 2 $C$ Fit} & 
\multicolumn{2}{c}{Doublet Method}  &  \multicolumn{2}{c}{Full Coverage\tablenotemark{b}} \\ 
\colhead{Ion}  &\colhead{2$+$2a\tablenotemark{c}} & \colhead{4a\tablenotemark{c}} & \colhead{2$+$2a} & \colhead{4a} & 
\colhead{2$+$2a} & \colhead{4a} & \colhead{2$+$2a} & \colhead{4a}}
\startdata
C IV   & 1.6  & 0.9  & 2.1  & 1.1 & 3.3 &  1.6\tablenotemark{d}  & 1.5 & 0.9 \\
N V  & $>$6.6  & 2.1  & $>$6.6  & 2.1  & $>$8.1 &  3.6 & 4.7 & 2.0 \\
O VI  & $>$14 & $>$13 & $>$14  & $>$13  & $>$14  & $>$13 & 9.8 & 9.5 \\
H I  & 7.5  & \nodata  & 5.0  &  \nodata & \nodata &  \nodata & 4.4\tablenotemark{e} & \nodata \\
\enddata
\tablenotetext{a}{All column densities in units of 10$^{14}$ cm$^{-2}$; 
for doublets, measurements are for the red member; lower limits are quoted for lines 
with regions having $\tau \geq$ 3}
\tablenotetext{b}{$C =$1 assumed everywhere.}
\tablenotetext{c}{Component 2$+$2a, $v =-$230 -- $-$315 km s$^{-1}$; Component 4a, $v =-$435 -- $-$535 km s$^{-1}$}
\tablenotetext{d}{Integrated over limited range due to infinities in solution.}
\tablenotetext{e}{Using Ly$\gamma$ line.}
\end{deluxetable*}

   An additional probe of the nuclear geometry is possible because
the Ly$\beta$ absorption lies on the high-velocity blue wing of the O\,{\sc vi}
BLR profile, at $v_{BLR} \approx -$2200 km s$^{-1}$, as discussed in \S 3.1 and illustrated 
in Figure 3.  Thus, it samples different BLR kinematics than most other absorption lines, 
which absorb BLR velocities coinciding with the absorption outflow velocities 
($v_{BLR} \approx -$300 km s$^{-1}$).
Therefore, comparison of the Ly$\beta$ emission-line covering factor and
those associated with other lines serves as a probe of the kinematic-spatial 
structure of the BLR.  
The potential affect of complex velocity-dependent structure in the BLR on 
the absorption covering factors was explored in \citet{sria99}.  
Since we assumed a priori that all lines share the same individual 
covering factors in the Lyman series fit in \S 4.2, a legitimate comparison requires obtaining 
independent constraints on the covering factor of the O\,{\sc vi} BLR blue wing by the 
Ly$\beta$ absorber.  This comes straightforwardly from the observed Ly$\gamma$ absorption 
line, since Ly$\gamma$ has no underlying line emission.
Using the result that $C^c =$1, the intrinsic optical depth ratio for 
Ly$\beta$ : Ly$\gamma$, and the observed Ly$\gamma$ profile, 
the absorption profile for Ly$\beta$ can be derived as a function of $C^l$ 
from equation 4.  Some illustrative results are shown in Figure 10. 
This shows, for example, that both an unocculted and fully occulted high velocity 
O\,{\sc vi} BLR are ruled out by the data.
We find that values ranging from 0.5 $\lesssim C^l \lesssim$ 0.8 are required to fit the 
majority of the Ly$\beta$ absorption in components 2 -- 2a.
This is similar to the emission line covering factor derived from the global fit to
the CNO doublets (Figure 5b).
Additionally, the Ly$\alpha$ $C^l$ profile can be derived independently for comparison with Ly$\beta$,
in a similar manner as for Ly$\beta$ (i.e., using $C^c =$1, the observed Ly$\gamma$ profile, 
and the Ly$\alpha$ : Ly$\gamma$ $\tau$ ratio).  The result is identical to the solution
to the combined Lyman lines in components 2 -- 2a shown in Figure 5a within uncertainties.
Therefore, the absorption covering factor of the O\,{\sc vi} BLR emission at 
$v_{BLR} \approx -$2200 km s$^{-1}$ is similar to $C^l$ at 
lower emission-line velocities (by the CNO doublets and Ly$\alpha$).
These results may provide constraints for models of the BLR, e.g., testing disk vs 
spherical geometries and outflow vs rotational kinematics for the BLR.

\begin{figure*}[location=t]
\includegraphics[angle=90,width=17cm]{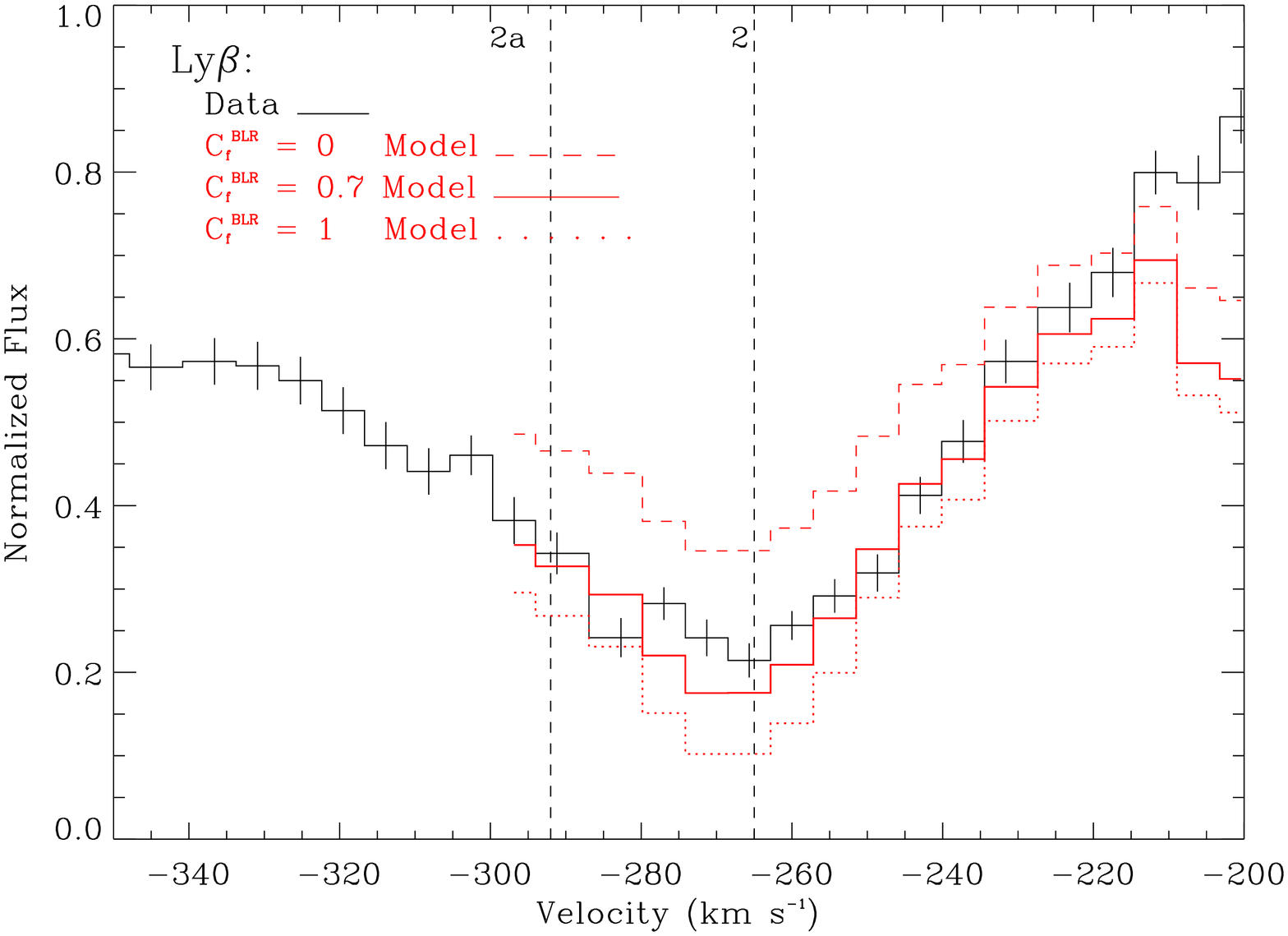}
\vspace*{0.1 in}
\caption{Constraints on the covering factor of the high blueshifted velocity O\,{\sc vi} BLR by
the Ly$\beta$ absorber.  The observed normalized Ly$\beta$ profile (black) is compared with 
models derived for three values of emission-line covering factor (red).  
Models were derived using the observed Ly$\gamma$ absorption, as described in the text.
Models with an unocculted (dashed) and fully occulted (dotted) BLR are ruled out.
An emission line covering factor of 0.7 (solid), consistent with the global fits, matches
most of the observed profile well. \label{fig10}}
\end{figure*}

\section{Summary}

     We have presented a study of the intrinsic UV absorption
in the Seyfert 1 galaxy Mrk 279 from an analysis of combined
long observations with {\it HST}/STIS and {\it FUSE}.
These spectra were obtained simultaneously in May 2003 as part
of an intensive multiwavelength observing campaign.

     We present a review of the standard technique for 
measuring intrinsic UV absorption parameters based
on individual doublet pairs, showing some key limitations of this
method:
1) It cannot treat multiple background emission sources. This 
introduces a potential error in the solution and misses 
important geometric constraints on the outflow.
2) Using synthetic absorption profiles, we show it systematically
underestimates the covering factor (and overestimates $\tau$)
in response to spectral noise.  The discrepancy in the solution 
is shown to be strongly dependent on absorption strength.

  To measure the UV absorption parameters in Mrk 279, we independently
fit two groups of lines:  the Lyman series 
lines and the combined CNO lithium-like doublets.
The doublet fitting involved a global fitting approach, which
assumes the same covering factors apply to all ions. 
By increasing the number of lines that are simultaneously fit, 
more complex and physically realistic models of the 
absorption-emission geometry can be explored.
Solutions for two different geometrical models, one assuming a single
covering factor for all background emission and the other separate 
covering factors for the continuum and emission lines, 
both give good statistical fits to the observed absorption.
However, several lines of evidence support the model with two covering
factors:
1) the independently fit Lyman lines and CNO doublets give 
similar solutions to the covering factors of both emission sources; 
2) the fits are consistent with absorbers that fully occult the 
continuum source and partially cover the emission lines,
consistent with the relative sizes of the emission sources; and 
3) observed variability in the Ly$\alpha$ absorption depth can be 
explained naturally by this model as a change in effective
covering factor resulting from a change in the relative strengths
of the emission components.

   Comparison of the traditional solutions  
based on individual doublets and the global-fit solutions shows the 
former exhibits much stronger velocity dependence.
This is seen as decreases in covering factor in the wings of individual 
kinematic components, and as peculiar fluctuations in both $\tau$
and $C$ in other regions of relatively weak absorption.
In light of the systematic errors shown to be inherent in the individual
doublet solution, we conclude some of these effects are likely artifacts
of the solution and should be interpreted with caution. 

   The covering factor solutions from our global fit constrains the relative line-of-sight
geometry of the absorbers and nuclear emission sources. 
The derived emission line covering factor, combined with the size
of the BLR, constrains the projected size of the absorber to be 
$\gtrsim$ 10 light days.  We utilize the coverage of the high velocity 
O\,{\sc vi} BLR by the Ly$\beta$ absorber to explore kinematic structure
in the BLR; we find no evidence for dependence of the absorber's BLR covering factor
on the BLR velocity.

\bigskip
    Support for this work was provided by NASA through grants HST-AR-9536, 
HST-GO-9688, and NAGS 12867 and through {\it Chandra} grant 04700532.
We thank D. Lindler and J. Valenti for their assistance in correcting the
STIS spectrum and C. Markwardt for making his software publicly available.
We also thank the referee P. Hall for comments that helped clarify and improve 
this study.

\section{Appendix}

     Here, we address how the expressions for covering factor and optical
depth from the doublet solution (equations 2 and 3) depend on
noise in the spectrum.  We have generated synthetic absorption profiles for 
doublet pairs that include random fluctuations to simulate spectral noise. 
We derive the $C$ -- $\tau$ solutions for these synthetic profiles to 
determine any trends in the solution as a function of noise level and absorption
strength. 

     Our synthetic profiles were derived in the following way.
The optical depth profiles were assumed to be Gaussian, parameterized by
the width ($\sigma$) and peak optical depth in the core of the blue line 
($\tau_{max}$).  The ``true" normalized absorption profiles for the lines 
are then derived from equation 1, with the covering factors set at a constant 
value across the profile for each doublet pair, and $\tau(v)_r = \tau(v)_b$/2 
at all velocities.  Thus, we have assumed in effect a single background emission source
to avoid the complications described in \S 3.1.  
For comparison with our study of Mrk 279, we have set 
the velocity resolution of our synthetic profiles to that of the STIS E140M 
grating, and the absorption width ($\sigma$ = 50 km s$^{-1}$) to be approximately 
consistent with intrinsic absorption components 2--2c.  To simulate spectral noise, 
we generated a normally distributed random number associated with each velocity 
bin in the synthetic spectra.  The noise level was normalized by selecting the 
desired S/N in the unabsorbed continuum and then weighting the noise by 
flux level, S/N$(v) \propto I(v)^{1/2}$, according to Poisson statistics.

  We generated profiles for a range of S/N, $C$ (real), and $\tau_{max}$.  
Here, we give some brief illustrative results, while reserving a full
analysis for a later study.
The left panels in Figure 11 show synthetic profiles for doublet pairs with 
$\tau_{max} =$ 2, 1, 0.5, and 0.25.  In all cases, $C =$0.8 and the noise 
level was normalized to be 3\% in the continuum.  The middle two panels show the 
corresponding covering factor and optical depth profiles derived directly from equations 2 and 3, 
with the actual values marked with dashed lines for comparison.
Errors in the doublet solution are immediately apparent. 
The covering factors are systematically underestimated, 
and the magnitude of error is strongly dependent on absorption strength.
This is seen both in the lower covering factors derived in the cores of
features with lower $\tau_{max}$ (means and standard deviations measured 
over the central 150 km s$^{-1}$ are printed in each plot), and in the
decrease in $C$ computed in the wings of each profile.

\begin{figure*}[location=center]
\centering
\includegraphics[angle=90,width=18cm]{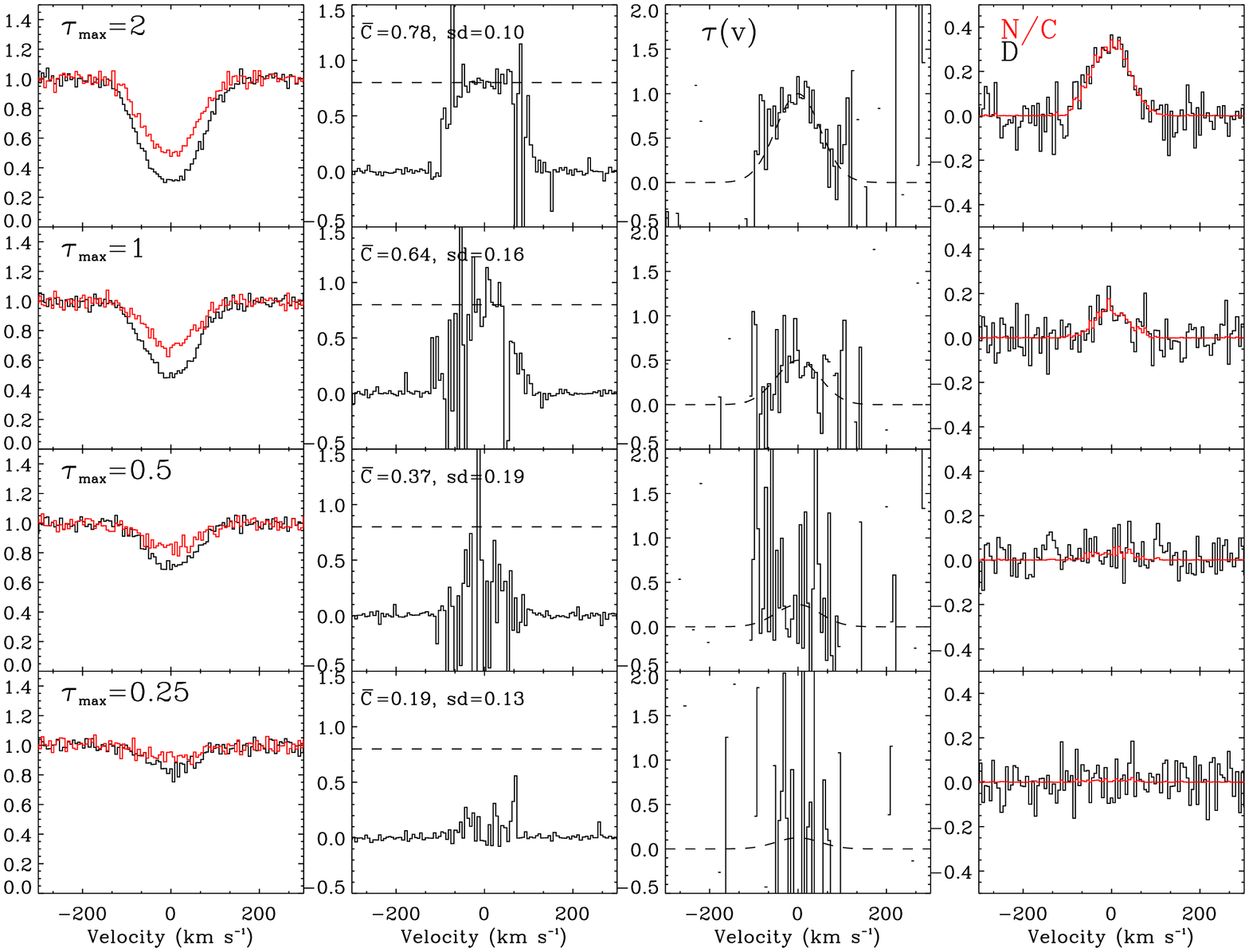}
\vspace*{0.35 in}
\caption{Systematic errors in the doublet solution resulting
from spectral noise.  Left panels show synthetic normalized doublet profiles (long
wavelength members in red) with simulated noise for a range of absorption strengths.  
The peak $\tau$ in the blue doublet member is given at the top of each panel.  Middle panels show
$C$ profiles derived from the doublet solution (equation 3); means and standard deviations 
(sd) measured in the central 150 km s$^{-1}$ are printed.  The actual constant
covering factor, $C =$0.8, is denoted with a dashed line for comparison.
Right panels show the value of the numerator ($N$) of the doublet equation divided by
$C$ (red), which would be identical with the denominator ($D$, black) for infinite S/N.
However, due to the non-linear form of $N$, $N / C$ drops below the noise level in $D$
for low $\tau$.  The result is a sytematic underestimation of $C$, with increasing
discrepancy for weaker absorption. \label{fig11}}
\end{figure*}

   These systematic errors are due to non-linear effects in the doublet equations.
This is seen most clearly by comparing the numerator, $N = (I_r - 1)^2$, and 
denominator, $D = I_b - 2 I_r + 1$, in the expression for covering factor (equation 2).
The right panels of Figure 11 show the values of $D$ and $N / C$ (in red), 
which would be identical in each velocity bin for infinite S/N.
Due to the forms of $N$ and $D$, these quantities have very
different dependences on noise; $\Delta N / N$ becomes much smaller than 
$\Delta D / D$ for weak absorption and, at sufficiently small $\tau$, 
$N$ is less than the noise level of $D$.
As $N$ decreases relative to the noise in $D$ for weaker absorption,
the probability that $0 \leq D \leq N / C$ becomes vanishingly small, and the average value of
$N / D$ becomes increasingly small.

   These errors could have pronounced effects on the interpretation of the outflow.
Each solution that underestimates $C$ overestimates $\tau$.
Thus, ionic column densities are systematically overestimated, with
increasing relative discrepancies in weaker doublets, leading 
to errors in determining the ionization structure and total gas in the
absorber via photoionization models.
Additionally, the errors in covering factor solutions will effect  
geometric inferences.
For example, due to the high-ionization state of AGN outflows, 
lower-ionization species appearing in UV spectra are generally weaker.
Thus, the increasing discrepancy in weaker absorption doublets may lead to 
the misinterpretation of ionic-dependent covering factors.
Also, weaker absorption in the wings of an absorption feature could lead to 
apparent velocity-dependent covering factors that are
instead due to optical depth variations, or at least exaggerate the effect.

\clearpage


\clearpage
\begin{thebibliography}{}
\bibitem[Arav et al.(2004)]{arav04}Arav, N., et al. 2004, \apj, in press 
\bibitem[Arav, Korista, \& de Kool (2002)]{arav02}Arav, N., Korista, K. T., \& de Kool, M. 2002, \apj, 566, 699
\bibitem[Arav et al.(1999)]{arav99}Arav, N., Becker, R. H., Laurent-Muehleisen, S. A., Gregg, M. D.,
White, R. L., Brotherton, M. S., \& de Kool, M. 1999, ApJ, 524, 566
\bibitem[Blandford \& Begelman (2004)]{blan04}Blandford, R, D., \&  Begelman, M. C. 2004, MNRAS, 349, 68
\bibitem[Blandford \& Begelman (1999)]{blan99}Blandford, R, D., \&  Begelman, M. C. 1999, MNRAS, 303, L1
\bibitem[Barlow \& Sargent (1997)]{barl97}Barlow, T. A., \& Sargent, W. L. W. 1997, \aj, 113, 136
\bibitem[Bevington (1969)]{bevi69}Bevington, P. R. 1969, Data Reduction and Error Analysis for the Physical Sciences, pp 242-245
\bibitem[Bohlin, Kickinson, \& Calzetti (2001)]{bohl01}Bohlin, R.~C., Dickinson, M.~E., \& Calzetti, D.\ 2001, \aj, 122, 2118
\bibitem[Cardelli, Clayton, \& Mathis (1989)]{card89}Cardelli, J. A., Clayton, G. C., \& Mathis, J. S. 1989, \apj, 345, 245
\bibitem[Cavaliere, Lapi, \& Menci (2002)]{cava02}Cavaliere, A., Lapi, A., \& Menci, N., 2002, ApJ, 581, L1
\bibitem[Cohen et al.(1995)]{cohe95}Cohen, M. H., Ogle, P. M., Tran, H. D., Vermeulen, R. C., Miller, J. S., Goodrich, R. W. \& Martel, A. R. 1995, ApJ, 448, L77
\bibitem[Costantini et al.(2004)]{cost04}Costantini, E., et al. 2004, \apj, in press 
\bibitem[Crenshaw, Kraemer, \& George (2003)]{cren03}Crenshaw, D. M., Kraemer, S. B., \& George, I. M. 2003, AARA, 41, 117
\bibitem[Crenshaw et al.(1999)]{cren99}Crenshaw, D. M., Kraemer, S. B., Boggess, A., Maran, S. P., Mushotzky, R. F., 
\& Wu, C.-C. 1999, \apj, 516, 750
\bibitem[de Kool, Korista, \& Arav (2002)]{deko02}de Kool, M., Korista, K. T., \& Arav, N. 2002, ApJ, 580, 54
\bibitem[de Kool et al.(2001)]{deko01}de Kool, M., Arav, N.,  Becker, R., Laurent-Muehleisen, S. A., White, R. L.,
Price, T., Gregg, M. D. 2001, ApJ, 548, 609
\bibitem[Gabel, Kraemer, \& Crenshaw (2004)]{gabe04}Gabel, J. R., Kraemer, S. B., \& Crenshaw, D. M. 2004, 
in AGN Physics with the Sloan Digital Sky Survey, eds. G. T. Richards \& P. B. Hall (San Francisco: ASP), p239
\bibitem[Gabel et al.(2003)]{gabe03}Gabel, J. R., et al. 2003, ApJ, 583, 178
\bibitem[Ganguly et al.(1999)]{gang99}Ganguly, R., Eracleous, M., Charlton, J. C., \& Churchill, C. W. 1999, \aj, 117, 2594
\bibitem[George et al.(1998)]{geor98}George, I. M., Turner, T. J., Netzer, H., Nandra, K., Mushotzky, R. F., \&
Yaqoob, T. 1998, \apjs, 114, 73
\bibitem[Goodrich \& Miller (1995)]{good95}Goodrich, R. W., \& Miller, J. S. 1995, \apj, 448, L73
\bibitem[Hall et al.(2003)]{hall03}Hall, P. B., Hutsemekers, D., Anderson, S. F., Brinkmann, J.,
Fan, X., Schneider, D. P., York, D. G. 2003, \apj, 593, 189
\bibitem[Hamann et al.(1997)]{hama97}Hamann, F., Barlow, T. A., Junkkarinen, V., \& Burbidge, E. M. 1997, \apj, 478, 80
\bibitem[Heap \& Brown (1997)]{heap97}Heap, S.~R.~\& Brown, T.~M.\ 1997, in The 1997 HST Calibration Workshop
with a New Generation of Instruments, ed. S. Casertano (Baltimore: STScI), 114
\bibitem[Kaastra et al.(2004)]{kaas04}Kaastra, J. S., et al. 2004, \apj, in press 
\bibitem[Kaspi et al.(2000)]{kasp00}Kaspi, S., Smith, P. S., Netzer, H., Maoz, D., Jannuzi, B. T., \& Giveon, U.
2000, ApJ, 533, 631
\bibitem[Kriss (2002)]{kris02}Kriss, G. A.  2002, in ASP Conf. Ser. 255, Mass Outflow in Active Galactic Nuclei: New 
Perspectives, ed. D. M. Crenshaw, S. B. Kraemer, \& I. M. George (San Francisco: ASP), 69
\bibitem[Kraemer et al.(2001)]{krae01}Kraemer, S. B., et al.  2001, \apj, 551, 671
\bibitem[Kraemer et al.(2002)]{krae02}Kraemer, S. B., Crenshaw, D. M., George, I. M., Netzer, H., Turner, T. J.,
\& Gabel, J. R.  2002, ApJ, 577, 98
\bibitem[Laor \& Netzer (1989)]{laor89}Laor, A. \& Netzer, H.  1989, MNRAS, 238, 897
\bibitem[Lindler \& Bowers (2000)]{lind00}Lindler, D., \& Bowers, C. 2000, BAAS, 197, 1202
\bibitem[Maoz et al.(1990)]{maoz90}Maoz et al. 1990, ApJ, 351, 75
\bibitem[Peterson et al.(1998)]{pete98}Peterson, B. M., Wanders, I., Bertram, R., Hunley, J. F., Pogge, R. W.,
Wagner, R. M. 1998, ApJ, 501, 82
\bibitem[Proga (2003)]{prog03}Proga, D.  2003, ApJ, 585, 406
\bibitem[Reynolds (1997)]{reyn97}Reynolds, C. S. 1997, MNRAS, 286, 513
\bibitem[Scannapieco \& Oh (2004)]{scan04}Scannapieco, E., \& Peng Ho, S., 2004, ApJ, submitted, astro-ph/0401087   
\bibitem[Scott et al.(2004)]{scot04}al. 2004, ApJS, 152, 1 (SK04)
\bibitem[Silk \& Rees (1998)]{silk98}Silk, J., \& Rees, M. J., 1998, A\&A 331, L1S
\bibitem[Srianand \& Shankaranarayanan (1999)]{sria99}Srianand, R., \& Shankaranarayanan, S., 1999, \apj, 518, 672
\bibitem[Wakker et al.(2001)]{wakk01}Wakker, B. P., Kalberla, P. M. W., van Woerden, H., de Boer, K. S., \& Putman, M. E. 2001,
ApJS, 136, 537
\bibitem[Wampler, Bergeron, \& Petitjean (1993)]{wamp93}Wampler, E. J., Bergeron, J., \& Petitjean, P. 1993, A\&A, 273, 15
\end{thebibliography}
\end{document}